\documentclass[a4paper,useAMS,usenatbib]{mn2e}

\usepackage[total={17.8cm,24.0cm},centering]{geometry}
\usepackage[flushleft]{threeparttable}
\usepackage{graphicx}
\usepackage{float}
\usepackage{times}
\usepackage{color}
\usepackage{subcaption}
\usepackage{url}
\usepackage{amsmath}
\usepackage{accents}
\usepackage{xtab,afterpage}
\usepackage{lipsum}  
 \usepackage{pdflscape}
\usepackage[dvipsnames]{xcolor}

\def\mcyl{$m_{\rm{cyl}}$}

\newcommand{\sauron}{{\texttt {SAURON}}}

\title[NGC~7457: A peculiar disc galaxy]
{NGC~7457: Evidence for merger-driven cylindrical rotation in disc galaxies}

\author[Molaeinezhad et al.]
{A. Molaeinezhad$^{1,2}$\thanks{Email: amolaei@iac.es},
	L. Zhu$^{3}$,
	J. Falc\'on-Barroso$^{1,2}$,
	G. van de Ven$^{4}$,
	J. M\'endez-Abreu$^{1,2}$,
	\newauthor
	M. Balcells$^{1,5}$,
	J. A. L. Aguerri$^{1,2}$,
	A. Vazdekis$^{1,2}$,
	H. G. Khosroshahi$^{6}$,
	R. F. Peletier$^{7}$ \\
	$^{1}$Instituto de Astrof\'isica de Canarias, E-38200, La Laguna, Spain\\
	$^{2}$Depto. Astrof\'isica, Universidad de La Laguna (ULL), E-38206 La Laguna, Tenerife, Spain \\
	$^{3}$Shanghai Astronomical Observatory, Chinese Academy of Sciences, Shanghai 200030, China\\
	$^{4}$Department of Astrophysics, University of Vienna, T\"urkenschanzstrasse 17, 1180 Vienna, Austria\\
	$^{5}$Isaac Newton Group of Telescopes, Apartado 321, 38700 Santa Cruz de La Palma, Canary Islands, Spain \\
	$^{6}$School of Astronomy, Institute for Research in Fundamental Sciences (IPM), PO Box 19395-5746 Tehran, Iran\\
	$^{7}$Kapteyn Astronomical Institute, University of Groningen, Postbus 800, 9700 AV Groningen, the Netherlands }

\begin{document}
 \date{Accepted 2019 June 24. Received 2019 June 17; in original form 2019 April 9}


\maketitle
\label{firstpage}

\begin{abstract}

We construct Schwarzschild orbit-based models of NGC~7457, known as a peculiar low-mass lenticular galaxy. Our best-fitting model successfully retrieves most of the unusual kinematics behaviours of this galaxy, in which, the orbital distribution of stars is dominated by warm and hot orbits. The reconstructed surface brightness of the hot component matches fairly well the photometric bulge and the reconstructed LOSVD map of this component shows clear rotation around the major photometric axis of the galaxy. In the absence of a dominant cold component, the outer part of our model is dominated by warm orbits, representing an exponential thick disc. Our orbital analysis also confirm the existence of a counter-rotating orbital substructure in the very centre, reported in previous observational studies. By comparing our model with a variety of simulation studies, and considering the stellar kinematics and populations properties of this galaxy, we suggest that the thick disc is most likely a dynamically heated structure, formed through the interactions and accretion of satellite(s) with near-polar initial inclination. We also suggest a merger-driven process as the most plausible scenario to explain the observed and dynamically-modelled properties of the bulge of NGC~7457. We conclude that both the high level of cylindrical rotation and unusually low velocity dispersion reported for the NGC~7457 have most-likely external origins. Therefore, NGC~7457 could be considered as a candidate for merger-driven cylindrical rotation in the absence of a strong bar in disc galaxies.

\end{abstract}


\section{Introduction}
\label{sec:intro}
As a prediction of the cold dark matter cosmological model \citep[$\Lambda$CDM;][]{spri2006}, hierarchical clustering  \citep{whit1997, stei2001} and major mergers \citep{toom1977} build elliptical galaxies and classical (elliptical-like) bulges of disc galaxies on a short time scale through the complete and violent collapse of the protogalactic clouds. In this schema, mergers get less common through the expansion of the universe and consequently the evolution of galaxies, which are getting now more and more isolated, gradually changing to a more secularly driven one under the influence of internal rather than external actors \citep{toom1977, lefe2000,cons2003,korm2010,atha2013,knap2013}. This slow evolution of galaxies has been introduced as "secular evolution" \citep{korm1982,korm2004, korm2008}. However, mergers and accretion must also occur along the life of galaxies, even in low-density environments, and such external processes are known to push disc galaxies along the Hubble sequence toward Sa/S0 \citep{ague2001,elic2006, elic2011}. Much observational and numerical work has been devoted to discriminating between internal and external evolutionary mechanisms \citep[for a review see][]{korm2013}. 
Bars in disc galaxies are considered as key drivers of the secular evolution by redistribution of the angular momentum, triggering of star formation and the morphological transformation of galaxies. During the secular evolution phase, bars can significantly alter the structure and kinematics of disc galaxies and develop a complex rotation pattern within the central parts of galaxies, called  "cylindrical rotation" \citep{korm1982}. This feature is defined as almost constant stellar rotation speed within the bulge of a galaxy in the direction perpendicular to the disc plane, such that $\delta v / \delta |z| \sim 0$, where $v$ is the line-of-sight velocity and $z$ is the distance from the disc plane in edge-on projection. These feature has been confirmed in several observational studies of nearby barred galaxies observed in the edge-on view \citep[e.g.][]{penc1981,korm1983s, bett1997, emse2001, marq2003, pere2009, will2011, mola2016,mola2017}, including our Milky Way \citep[see][and references therein]{barb2018}. N-body simulations of barred galaxies confirm the link between the presence of a bar and a tendency to rotate on cylinders for stars within the bar, as a natural consequence of the bar buckling processes \citep[see][]{comb1981,mart2004, atha2003,bure2005,atha2005}. In this scenario, after the bar develops, it experiences a period of buckling instability that thickens it in the axial direction \citep[e.g.][]{comb1981}, giving birth to a boxy or peanut-shape bulgy structure in the central parts of the bar, termed boxy/peanut bulge (B/P, hereafter). From the galactic-dynamics point of view, bar formation makes the orbits in the bar more eccentric and more boxy and aligns their principal axes without greatly affecting the motion perpendicular to the plane; consequently, the cylindrical rotation follows naturally \citep{bure1997}. As demonstrated in numerous N-body simulation studies of barred galaxies, the resulting velocity field due to this orbital reconfiguration is compatible with the observed cylindrical rotation in B/P bulges \citep[see][and references therein]{mola2016}.

Cylindrical rotation was objectively-quantified by \citet{mola2016} for a sample of nearby early-to-intermediate disc galaxies with high-quality \sauron\ 2D spectroscopy. Their findings support that high levels of cylindrical rotation is a common feature of strongly-barred disc galaxies. They also note that NGC~7457 is an exception, in the sense that the level of cylindrical rotation is as high as that found in B/P bulges in their sample, while the photometric and kinematic properties of this galaxy suggest there is no bar present. In fact, the stellar line-of-sight velocity map of this galaxy, within the bulge dominated region, is distinctly different to that of an unbarred galaxies (classical bulge), in which the minor-axis rotation velocity is naturally expected to decrease along the vertical direction \citep[see Fig. 3 of ][for a comparison between the line-of-sight velocity maps and profiles of two representative cylindrically rotating and classical bulges]{mola2016}. This finding may pose a challenge to the generally-accepted belief that high level of cylindrical rotation traces the presence of a bar. It also brings up the question that, in absence of a strong bar, what mechanism can create high cylindrical rotation?

\begin{table}
\caption {Basic properties and characteristics of NGC~7457}
\label{tab:basic-info}
\begin{center}
\begin{tabular}{lll}
\hline
Parameter 	                                 				 & Value 			                  & Source			                                \\
\hline
R.A. (J2000)  (\textit{h:m:s})								 & 23:00:59.9				 & (1) 							\\
Dec. (J2000) (\textit{d:m:s}) 								& +30:08:42				    & (1) 							\\
m$_{B}$	($mag$)										 & 11.86					        & (2)          \\
m$_{I}$	($mag$)				                           & 9.45					          & (3)           \\
Distance (\ensuremath{\mathrm{Mpc}}) 				                      & 12.9					          & (4) 					\\
Morphological type 				                      & SA(rs)0-?				     & (2) 		\\
PA (N–E) ($deg.$)				                  & 35 						             & (5) 		         \\
Inclination ($deg.$)			                   & 74 						          & (6)                		 \\
Effective radius  (\ensuremath{\mathrm{arcsec}})			                   & 32 						          & (2)                		 \\
Scale (kpc/arcsec; H$_{0}$=70 \ensuremath{\mathrm{km\,s}^{-1}} \ensuremath{\mathrm{Mpc}^{-1}})    & 0.08      & (1)							 \\
Photometric bulge radii ($x_B$, $z_B$)       (\ensuremath{\mathrm{arcsec}})            & 11.0,9.5     	  & (5)                  		\\
M/L (Using HST/WFPC2 in the I--band)                                         & 1.86    	  & (3)                  		\\
V/$\sigma$		  (within 1\ensuremath{R_\mathrm{e}})		                                               & 0.62		& (7)                           \\
log(M$_{*}$) (\ensuremath{M_{\sun}})		               & 10.13					         & (4)				\\
$\lambda_{Re}$	(within 1\ensuremath{R_\mathrm{e}})	    & 0.570	                          & (7)                      \\
M$_{BH}$	(\ensuremath{M_{\sun}})			            & 3.5 $\times$ 10$^{6}$  	  	       & (8)				 \\
\hline
\end{tabular}
\end{center}
\begin{flushleft}
	\small  1: NED; 2: \citet{deva1991} (RC3); 3: \citet{capp2006}; 4: \citet{alab2017}; 5: \citet{mola2016}; 6: \citet{capp2013}; 7: \citet{capp2007}; 8: \citet{gebh2003}
\end{flushleft}

\end{table}

\begin{figure*}
	\captionsetup[subfigure]{labelformat=empty}
	\centering
	\begin{subfigure}{0.55\textwidth}
		\centering
		\includegraphics[width=1.0\textwidth]{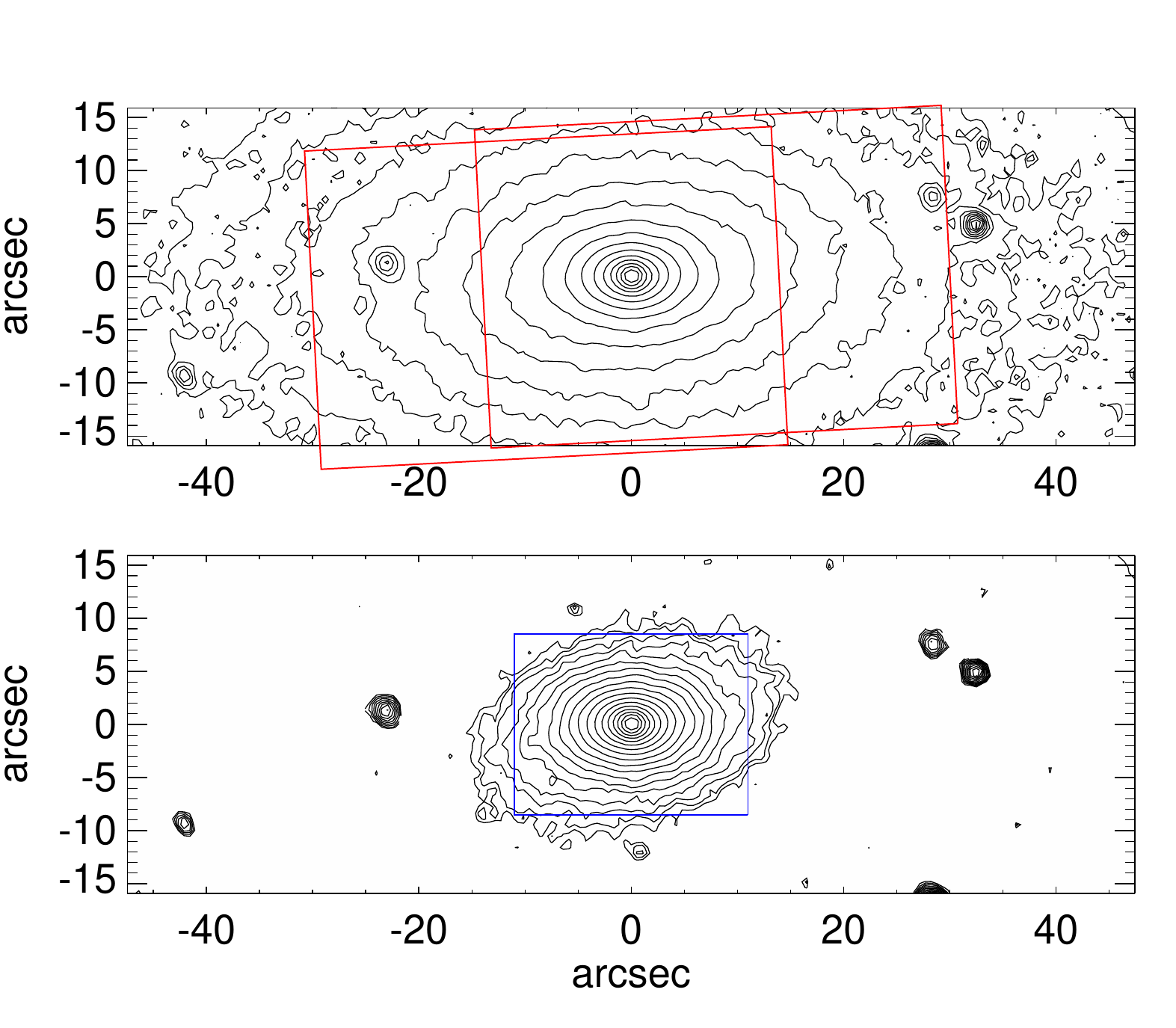}
		\label{fig:sch-decom4-i90}
	\end{subfigure}%
	\begin{subfigure}{0.39\textwidth}
		\centering
		\includegraphics[width=1.0\textwidth]{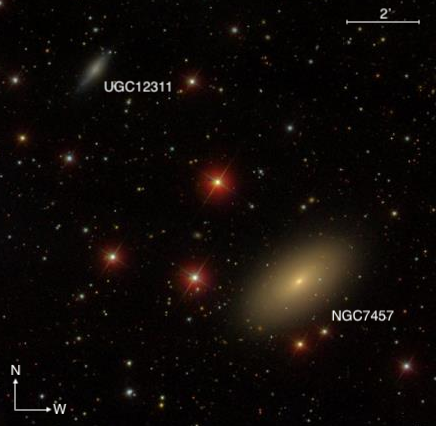}
		\label{fig:sch-decom4-i0}
	\end{subfigure}
	\centering
	\centering
	\caption{Left-top panel: Isophotes of the r--band SDSS image of NGC7457. The two \sauron\ pointings are shown on top of this panel. Left-bottom panel: Isophotes of the r--band image, after subtracting the best exponential disc from the observed image. The bulge analysis window is shown as a blue rectangle on top this panel. Left panel: The SDSS gri color composite image (provided by the SDSS SkyServer ) of NGC~7457 ($13.4 ^{'} \times 13.4^{'}$) and its companion galaxy, UGC~12311. North is up, and East to the left.
		\label{fig:7457morph}}
\end{figure*}

NGC~7457 is a highly inclined S0 galaxy \citep[inclination $i=74^{o}$;][]{capp2013}, originally selected from the well-characterised sample of \citet{balc1994} of inclined early-type disc galaxies. This galaxy is one of the nearest S0 \citep{deva1991} with a distance of 12.9 \ensuremath{\mathrm{Mpc}} \citep{alab2017} and it resides in one the 3 subgroups of filamentary group NGC~7331 \citep{ludw2012}. It is accompanied by UGC 12311, a spindle-shaped galaxy at a distance of 5.7\arcmin\ from the centre of NGC 7457 (see right panel of Fig.~\ref{fig:7457morph}). The basic properties of  NGC~7457 are presented in Table~\ref{tab:basic-info}. 
NGC~7457 is peculiar in several of its properties. Previous studies proposed a very small bar-like distortion in the centre \citep{mich1994}, a counter-rotating core \citep{silc2002}, a chemically distinct nuclei ($r<1\farcs5$) younger than the surrounding bulge \citep{silc2002} associated with AGN \citep{gebh2003}, a significantly high level of cylindrical rotation and unusually low measured velocity dispersion in the bulge \citep{silc2002, emse2004, mola2016} and a possible disc-like bulge on the basis of observed deviations from the Faber–Jackson $L$--$\sigma$ relation \citep[e.g.][]{korm1993,pink2003}. Both chemodynamical and photometric studies of NGC~7457 bulge, confirm the bulge of NGC 7457 is young, with a mean age of the stellar population of 6--7 \ensuremath{\mathrm{Gyr}} \citep{andr1995,silc2002}.

The luminosity profile of the NGC~7457 bulge has also been a matter of debate. While \citet{burs1979} classified this galaxy as a disc-dominated system with $D/B=1.6$, \citet{andr1995} argued the slope of luminosity profile for bulges of this galaxy is much steeper ($n_{Sersic}=6 $) than one would expect for a disc-dominated 
galaxy, even steeper than namely normal lenticulars and of pure ellipticals. Recently, \citet{erwi2015} performed a detailed morphological and isophotal analysis of NGC~7457 using archival V--band images from JKT and INT telescopes, combined with the HST WFPC2 F555W images for the central regions ($r < 6\arcsec$). They claim that, the isophotal structure of this galaxy at both inner and outer regions could be very well described with a round classical bulge embedded within a highly inclined disc, with no evidence in favour of any bar-like stellar structure in the isophotes and surface brightness (SB, hereafter) profiles. 

From a different point of view, \citet{harg2011} performed a detailed chromodynamical analysis 
of globular clusters (GCs) in this galaxy, as a proxy for galaxy formation models and reported a noticeably elliptical spatial distribution for GCs. They conclude that the formation scenario for this galaxy is most-likely through a merger event involving galaxies with unequal masses. This is in agreement with \citet{silc2002} who also proposed a a merger-driven evolution in this galaxy. \citet{zana2018} in contrast, using the same approach, found that NGC~7457 has a disc dominated GC population. They claimed that, considering the estimated halo assembly epoch of this galaxy, along with the low density environment of this system and discy distribution of its GCs, a secular evolution scenario is the best mechanism to explain the observed properties of this system.

Overall, these puzzling contradictions suggest that NGC~7457 could not be consider as a typical S0 and most likely this system has a complex evolutionary history. Our suggestion is that, cylindrical rotation as a key dynamical behaviour of the inner regions of this  galaxy may play a crucial role in this context and answering the question that "What is the origin of such high level of cylindrical rotation in bulge of NGC~7457?" could offer new insights into our understanding of the formation and evolution of this peculiar galaxy. It is worth noting that, most of the previous studies on this galaxy suffer from the uncertainties due to inconsistencies between the morphologically-- and kinematically--based decompositions approaches and the way that the outcomes of these two approaches are interpreted and linked to each other.

Schwarzschild orbit-based dynamical modelling \citep[see][and references therein]{zhu2018} has been widely used to uncover the internal dynamics of early-type galaxies. It builds galactic models by weighting the orbits generated in a gravitational potential, makes it possible to dynamically decompose the galaxies into different orbital/dynamical components \citep[e.g.][]{vand2008}. Recently, \citet{zhu2018} has extended the application of this approach to late-type galaxies. In this method, the dynamically decomposed components are not necessary to match the morphologically decomposed components and therefore does not suffer from the uncertainties due to the model-dependent morphological decompositions \citep[see][]{tabo2017, mend2018}. 

In this paper, as the third paper of a series aimed to understand the nature and evolutionary processes of bulges in highly-inclined disc galaxies, we focus on the origin of kinematics peculiarities, reported in previous studies for this galaxy, specially, the high level of cylindrical rotation and low velocity dispersion in the NGC~7457 bulge, based on a detailed dynamical study of this system, derived from the orbit-superposition Schwarzschild models. In Section~\ref{sec:data}, we present our data and give a summary of optical properties, kinematics and stellar population analyses of the this system. In Section~\ref{sec:schw-method}, we present our results of the Schwarzschild orbit--superposition modelling of NGC~7457 and show the orbital decompositions. Section~\ref{sec:discussion} gives the discussion and we conclude in Section~\ref{sec:conclusion}.

\begin{figure*}
	\centering
	\includegraphics[width=0.92\textwidth]{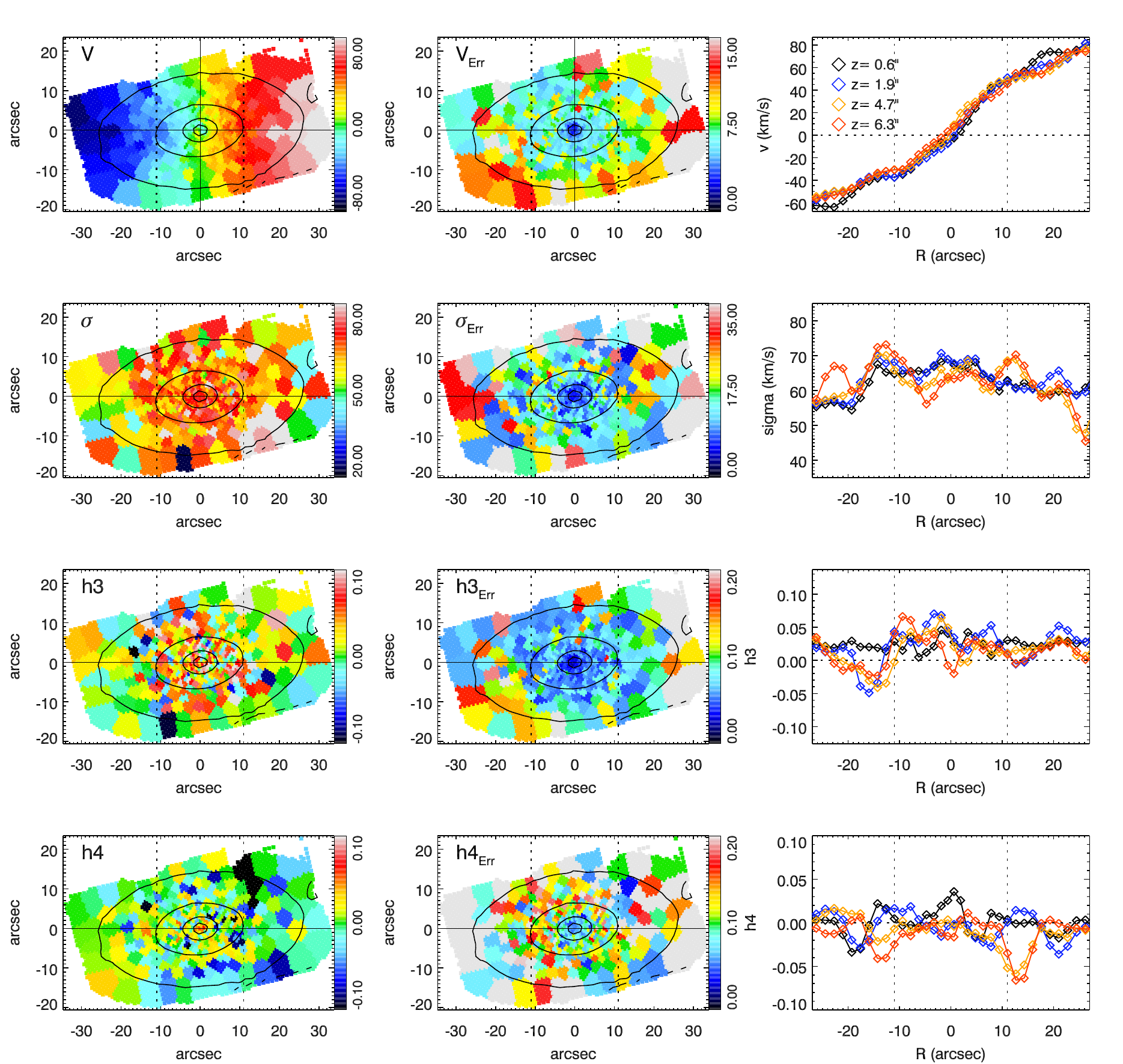}
	\caption{SAURON stellar kinematics maps, related error maps and profiles along cuts parallel to the major axis at four representative different 
heights from the disc plane. The vertical dashed lines show the bulge boundary along the major axis ($x_B$). Note that, all maps are rotated so that for each galaxy the x-axis are along the kinematic major axis of galaxy.}
	\label{fig:sauron_kin}
\end{figure*}

\section{Data \& Measurements}
\label{sec:data}
 The IFU observations of NGC~7457 has been described in detail in \citet{mola2016}. Briefly, the spectroscopic observations of this galaxy, as part of a project to study the vertical kinematics and population structure of bulges in highly inclined disc galaxies \citep[see][and references therein]{mola2016} were carried out between October 1999 and 2011 with the \sauron\  integral-field spectrograph \citep{baco2001} attached to the 4.2--m William Herschel Telescope (WHT) of the Observatorio del Roque de los Muchachos at La Palma, Spain. Considering the main aim of that project, the observed field of view (FoV) and spatial resolution for the survey has been considered to cover 1$R_{e}$ of this galaxy. For NGC~7457, two overlapped \sauron\  pointings give a $18\arcsec  \times 36\arcsec$ field-of-view (FoV) of the central region of this galaxy (see left-top panel of Fig.~\ref{fig:7457morph}). The stellar kinematics data used here is as the one presented in \citet{mola2016}. The kinematics maps for NGC~7457 are presented in Fig.~\ref{fig:sauron_kin}. In this figure, we present the stellar velocity, $\sigma$, h$_{3}$ and h$_{4}$ maps (left panels) and related error maps (middle panels). We also present the radial profiles for these quantities along the major axis of the galaxy at four different heights from the mid-plane (right panels), within the photometric bulge analysis window. Note that, following the same approach presented in \citet{mola2016}, the extent of the bulge in the radial ($x$) and vertical ($z$) directions has been evaluated by integrating the residual images along the minor (or major) axis, after subtracting the best exponential disc model from the r--band image of NGC~7457 (taken from the Sloan Digital Sky survey DR10 \citep{ahn2014}). The $x_B$ and $z_B$ represent the region in which more than 90 per cent of light comes from the bulge. This region is marked with a blue rectangle in the left-bottom panel of Fig.~\ref{fig:7457morph}, where we present the isophotes of the residual image, obtained after subtracting the best fitted disc from the original image.

\subsection{Kinematics analysis}

The velocity map of NGC~7457 clearly exhibits a cylindrical rotation pattern within the photometric bulge region. The velocity profiles are remarkably parallel at different heights from the mid-plane, even beyond the photometric bulge window. 
In \citet{mola2016}, we introduced a numerical metric for cylindrical rotation (\mcyl), optimised for 2D kinematic data to measure the importance of this feature.  \mcyl\ is generally a value between +1  (indicates pure cylindrical rotation) and 0 (where there is no sign of cylindrical rotation). Following this approach, we found a value of \mcyl\ = 0.83 $ \pm$ 0.06 within the photometric bulge area of NGC~7457 and \mcyl\ = 0.71 $ \pm$ 0.13 over the entire field of view of our \sauron\ velocity map. This value is remarkably high, even higher than that measured for strong B/P bulges in side-on bars with perfectly edge-on orientation.  Given the widely accepted link between high level of cylindrical rotation and presence of bars, it is crucial to investigate this galaxy from this aspect.

The complexity of various non-axisymmetric features found within disc galaxies such as bars and spiral arms are considered a key to the understanding of the evolutionary path of these systems. In edge-on, or highly inclined disc galaxies, bars are most easily (photometrically) recognised by the prominent bi-lobed, boxy or X-like shape formed by the stellar material above the disc plane, via the vertical buckling instability of the bar \citep[e.g.][]{kuij1995,bure1999,mart2004}. Meanwhile, bars oriented exactly parallel to the line-of-sight appear almost spherical and thus difficult to identify \citep[see][]{bure2004,atha2015}. 

NGC~7457 has been classified as a SA0-(rs) \citep{deva1991} galaxy with no observational evidence supporting the presence of a B/P bulge or an strong bar in the photometric data. \citet{mich1994} have reported the possible presence of a small bar-like distortion, based on isophotal analysis, however, more recent analyses \citep[e.g.][]{balc2007, fish2008, fish2010, korm2011,erwi2015} clearly identify a round bulge embedded within an inclined disc for NGC~7457, with no sign of a discy pseudo-bulge or even a small bar-like structure. In fact, the transition to identifying the round bulge comes from using NIR HST imaging which helped with spatial resolution and weaker dust obscuration \citep{balc2007}. From kinematic point of view, bars in disc galaxies exhibit specific signatures. \citet{bure2005} suggested, using N-body simulations, a number of bar diagnostics in highly inclined systems: (1) a double-hump rotation curve; (2) a broad central velocity dispersion peak with a plateau (and possibly a secondary maximum) at moderate radii and (3) an h3 profile correlated with velocity over the projected bar length. Later, all these diagnostics have been successfully expanded beyond the disc plane by \citet{lann2016}. \citet{mola2016} exploited this approach on 12 mid- to highly inclined galaxies, observed with \sauron\ IFU, and suggested that the correlation between the stellar line-of-sight velocity (V) and the h$_{3}$ Gauss--Hermite parameter is a very reliable bar diagnostic tool. As noted in that work, the success of this method is in the ability to unveil the hidden bar, even where the bar is not clearly visible in photometric data, due to its orientation or strength. As discussed in \citet{mola2016}, the V-h$_{3}$ correlation is negative or null over the entire field of NGC~7457, as expected for an unbarred galaxies. Therefore, from both photometric and kinematics points of view, there is no evidence in favour of a bar-like structure in this galaxy.

As shown in the second row of Figure~\ref{fig:sauron_kin}, NGC~7457 is a case of a low-dispersion galaxy ($<\sigma_{bulge}>$=71.3 $\pm$ 3.5 \ensuremath{\mathrm{kms^{-1}}}). Given the significant deviation of this galaxy from the $L-\sigma_{0}$ relation of S0–Sbc galaxies, NGC 7457 has an unusually low measured velocity dispersion. This matter is still under debate. While some authors consider this galaxy as a peculiar case \citep[e.g.][]{korm1983,dres1983}, \citet{silc2002} interpret it as due to a possible counterrotating component that weaken and suppress the observed rotation velocity along the line-of-sight. However, except the kinematically decoupled core (KDC, hereafter), observed in the centre of NGC~7457 \citep[e.g.][]{emse2004}, no observational evidence in favour of a counter-rotating disc in outer region of this galaxy has been reported so far. Indeed, more chemo-dynamical studies are necessary to confirm this scenario. We investigate this in Section~\ref{sec:schw-method}, where we present our Schwarzschild dynamical model of this galaxy.

\subsection{Stellar population analysis}
\label{sec:ssp}

For NGC~7457, we have measured a set of line-strength indices available within the wavelength range provided by \sauron\ , (e.g. $H\beta_{o}$, Fe5015 and Mg $b$) to estimate the stellar population properties. These line indices have been measure in the recently defined Line Index System (LIS) LIS-8.4 \AA\ \citep{vazd2010}. Our stellar population analysis are based on $\alpha$–MILES stellar population models \citep{vazd2015} and our method has been described in detail in \citet{mola2017}. Briefly, $\alpha$–MILES stellar population models have been computed over a wide range of ages and metallicities at both solar scale and for $[\alpha/Fe] = +0.4$. To cover the entire parameter space,  for both scaled-solar and $\alpha$--enhanced SSP models, we produce an interpolated grid of model line-strengths for different ages and metallicities. Then, we interpolate/linearly-extrapolate these scaled-solar and $\alpha$--enhanced grids to cover the $[Mg/Fe]$ range from -0.1 to +0.8 $dex$. Finally, the SSP-equivalent age, metallicity, and [Mg/Fe] for each spectrum is determined by the least linear $\chi^{2}$ fitting, and uncertainties of the best-fit parameters are determined via bootstrapping. 

\begin{figure*}
	\centering
	\includegraphics[width=0.98\textwidth]{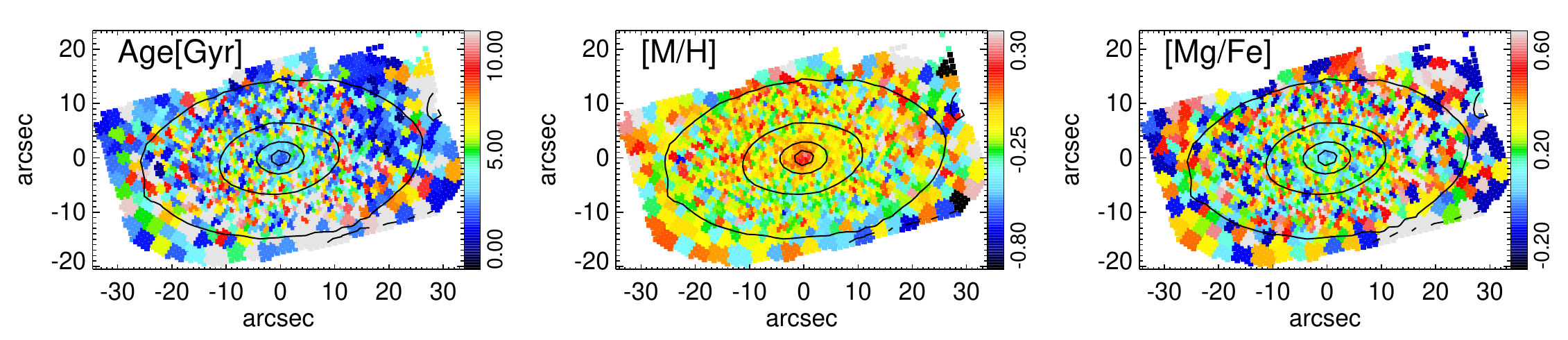}
	\caption{Computed SSP-equivalent maps of age, metallicity and [Mg/Fe] for NGC~7457. Overlaid on each image is the r--band isophotal contours, spaced by 2.0 magnitudes.}
	\label{fig:ssp-maps}
\end{figure*}

In Fig.~\ref{fig:ssp-maps} we present the SSP-equivalent age, metallicity, and [Mg/Fe] maps of NGC~7457 over the entire \sauron\ field. The stellar population over the entire field, covering the bulge  is remarkably young ($<age>=3.4 \pm 1.8$ \ensuremath{\mathrm{Gyr}}) and slightly older moving outwards while the very central region ($r< 2\farcs5$) is heavily dominated by 2-3 Gyr stellar populations. It is consistent with the central chemically distinct young stellar population reported for this galaxy by \citet{silc2002}, \citet{moor2006} and that from multi-band photometric studies \citep{pele1999}. The presence of such young populations of stars in the central regions of disc galaxies is mostly accompanied by large amounts of dust, while in case of NGC~7457 no nuclear dust feature was observed \citep{pele1999}.  

As shown in the middle panel of Fig.~\ref{fig:ssp-maps}, this young central region ($r< 2\farcs5$) also shows a sharp jump in the metallicity ($<[M/H]>=0.24 \pm 0.11$). The metallicities are then around solar values ($<[M/H]>=0.04 \pm 0.1$) in the surrounding regions inside a zone of radius $\approx$ 10\arcsec\ (20\arcsec) along the minor (major) axis and sub-solar values afterwards. 

The right panel of  Fig.~\ref{fig:ssp-maps}, illustrates a super-solar [Mg/Fe] in the bulge-dominated region of NGC~7457 ($<[Mg/Fe]>=0.35 \pm 0.19$), slightly lower within the central 2.5\arcsec. The [Mg/Fe] parameter is generally considered as a "chemical clock" to assess the timescale of star-formation for the bulk of a stellar population \citep[e.g.][]{thom2011,vazd2015}. It is worth noting that, \citet{mola2017} reported a strong difference in the behaviour of vertical [Mg/Fe]  gradients in bulges of barred and unbarred systems. In their sample of 38 mid to highly inclined nearby disc galaxies, the vertical [Mg/Fe] gradients within the bulge area of barred galaxies are mostly positive, while for unbarred galaxies the profile is almost flat or negative. While the [Mg/Fe] gradient for NGC~7457 is very similar to that of bulges in unbarred galaxies (excluding the very central region), the relatively young population of stars is not what one may expect from a classical bulge. This contradiction will be discussed in detail in Section~\ref{sec:discussion}.

\section{Schwarzschild models and orbital decomposition}
\label{sec:schw-method}
Understanding the distribution of stellar orbits  in a galaxy plays an important role in drawing a more realistic picture of its formation and evolution path and better understanding the observed properties. Schwarzschild’s (1979) orbit-superposition method is a powerful dynamical modelling technique that builds galactic models, using a representative library of stellar orbits in a gravitational potential and weighting them to reconstruct the observed surface brightness and kinematics of galaxies. 

\begin{table}
\caption {The best-fitting parameters of the triaxial Schwarzschild models for NGC~7457.}
\label{tab:schw-param}
\begin{center}
\begin{tabular}{ll}
\hline
Parameter 	                                 & Value \\
\hline
M$_{BH}$ (fixed parameter)								& 3.5 $\times$ 10$^{6}$ M$_\odot\ $ \\
Stellar mass-to-light ratio $\Upsilon_{*}$						& 2.29$_{-0.33}^{+0.15}$	\\
inclination angle $\vartheta$								& 73$^{o}$ $\pm$3 	\\
intrinsic shape of the model, $\bar{q}$ ($R_{e}$)				& 0.46$_{-0.03}^{+0.02}$ 	\\
intrinsic shape of the model, $\bar{p}$ ($R_{e}$)				& 0.77$_{-0.02}^{+0.02}$	\\
intrinsic shape of the model, $\bar{u}$ ($R_{e}$)				& 0.99$_{-0.01}^{+0.01}$	\\
DM mass, $M_{dm}$ ($ \leq R_{e}$)							& 0.46 $_{-0.21}^{+0.07}$ [$\times$ 10$^{10} M_\odot\ $]	\\
log($M_{200}/M_{*}$)												& 3.5$_{-0.51}^{+0.32}$	\\
\hline
\end{tabular}
\end{center}
\end{table}

\subsection{Schwarzschild orbit-based dynamical modelling}
\label{subsec:sch-dyn-modl}

Schwarzschild’s (1979) orbit-superposition technique has been described in detail in \citet{zhu2018}. Briefly and practically, the first step is to construct the stellar mass distribution of the 
gravitational potential, needed to build the triaxial Schwarzschild models. For this, a photometric model is constructed from the galaxy's image by 
an axisymmetric 2D Multiple Gaussian Expansion (MGE) model \citep{emse1994,capp2002}. After assuming the space 
orientation of the galaxy, described by three viewing angles ($\theta$, $\phi$, $\psi$), the 2D axisymmetric MGE flux could be de--projected to a 3D triaxial MGE luminosity density, in which the intrinsic shape of the 3D triaxial Gaussian components could be express by a set of three viewing 
angles ($q$,$p$,$u$), where qualitatively, $q$ represents the short-to-long axis ratio, $p$ is the intermediate to long axis ratio, and $u$ represents the ratio between the intrinsic and projected MGE Gaussian widths along the major axis of the galaxy. Assuming a constant stellar mass-to-light ratio $\Upsilon_{*}$ as a free parameter of the model and multiplying that to the  
reconstructed 3D luminosity, we obtain the intrinsic mass density of stars. For the DM distribution, a spherical NFW \citep{nava1996}
 halo is adopted with two free parameters: the DM concentration $c$ and the virial mass $M_{200}$, in which the concentration parameter could be 
fixed following the known relation with $M_{200}$ from cosmological simulations \citep[e.g.][]{dutt2014}. The problem is then reduced to 5 free parameters to 
build the models, namely $\Upsilon_{*}$, $q$, $p$, $u$ and $M_{200}$. A central BH mass is the last part that should be included to generate the gravitational potential that we fix it with $M_{BH} = 3.5 \times 10^6 M_{\odot}$ \citep{gebh2003}.  

\begin{figure*}
	\centering
	\includegraphics[width=1.0\textwidth]{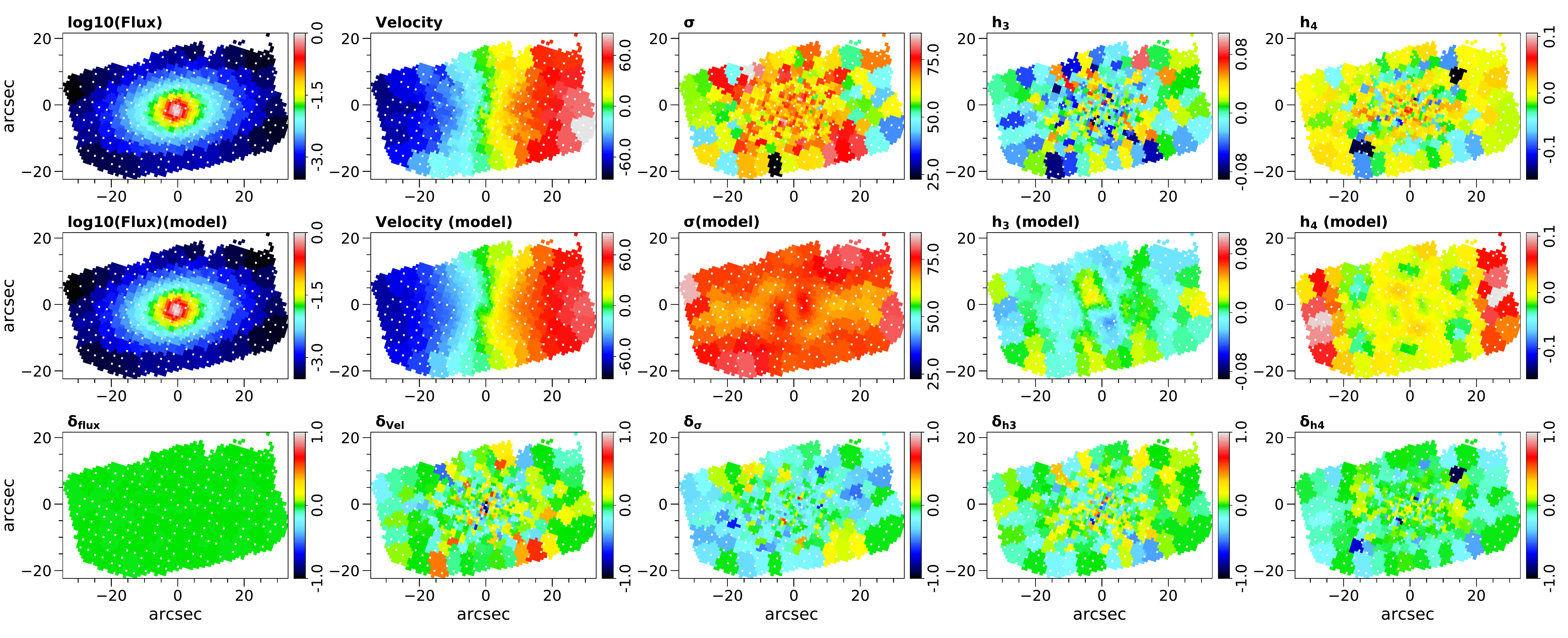}
	\caption{The kinematics maps for NGC~7457. The Top row demonstrates the observed surface brightness and kinematics maps. The middle row shows the best--fitting kinematics model and reconstructed surface brightness map. The bottom row illustrates the residual maps, we defined as the difference between the true (measured) and the reconstructed values, divided by the measured error per each Voronoi bin. Note that, all maps are rotated so that for each galaxy the x-axis are along the kinematic major axis of galaxy.}
	\label{fig:schw-kin-maps}
\end{figure*}

The next step to construct the triaxial Schwarzschild models is to generate a representative library of orbits that admit three exact analytical and isolating integrals of motion $E$, $I2$ and $I3$ in this gravitational potential \citep[see][for more details]{vand2008}. Orbits are characterised by the time-averaged radius, represents the size of each orbit and circularity parameter $\lambda_{z}$ that describes the angular momentum of each orbit in the given gravitational potential. In this context, $\lambda_{z}=1$ correspond to circular orbits,  $\lambda_{z}=0$ correspond to radial and box orbits, and counter--rotations orbits are identified with their negative circularities ($\lambda_{z} < 0$). 

The orbits weights, described by the probability density of orbit weights, $\rho(circularity, radius)$ are then determined by 
simultaneously fitting the orbit-superposition models to the observed surface brightness and stellar kinematics of the galaxy. Finally, to perform 
the orbital decomposition of the best fitted triaxial Schwarzschild model, a set of cuts is defined to decompose the orbits in the best-fitting model into four classes of hot, warm, cold and counter-rotating (CR) components.

For NGC~7457, we use our \sauron\ stellar kinematics maps, introduced in Section \ref{sec:data} to constrain the models. However, considering the higher S/N ration of our \sauron\ IFU data comparing to that of \citet{zhu2018}, we can also benefit from $h_{3}$ and $h_{4}$ maps to put more constrains on the models. Considering the wavelength range of our IFU data (4800--5380 \AA), the 2D flux distribution to construct the 3D stellar mass distribution is taken from the SDSS images in r-band.

 Fig. \ref{fig:schw-kin-maps} illustrates the reconstructed stellar kinematics maps for the best-fitting model for NGC~7457 (second row), well matched to the observed (inputs) surface brightness and kinematic maps (first row). In Table~\ref{tab:schw-param} we summarised the parameters of the best-fitting models. Errors of the best-fitting parameters have been calculated using the models within 1$\sigma$ confidence intervals. 

\begin{figure}
	\centering
	\includegraphics[width=0.5\textwidth]{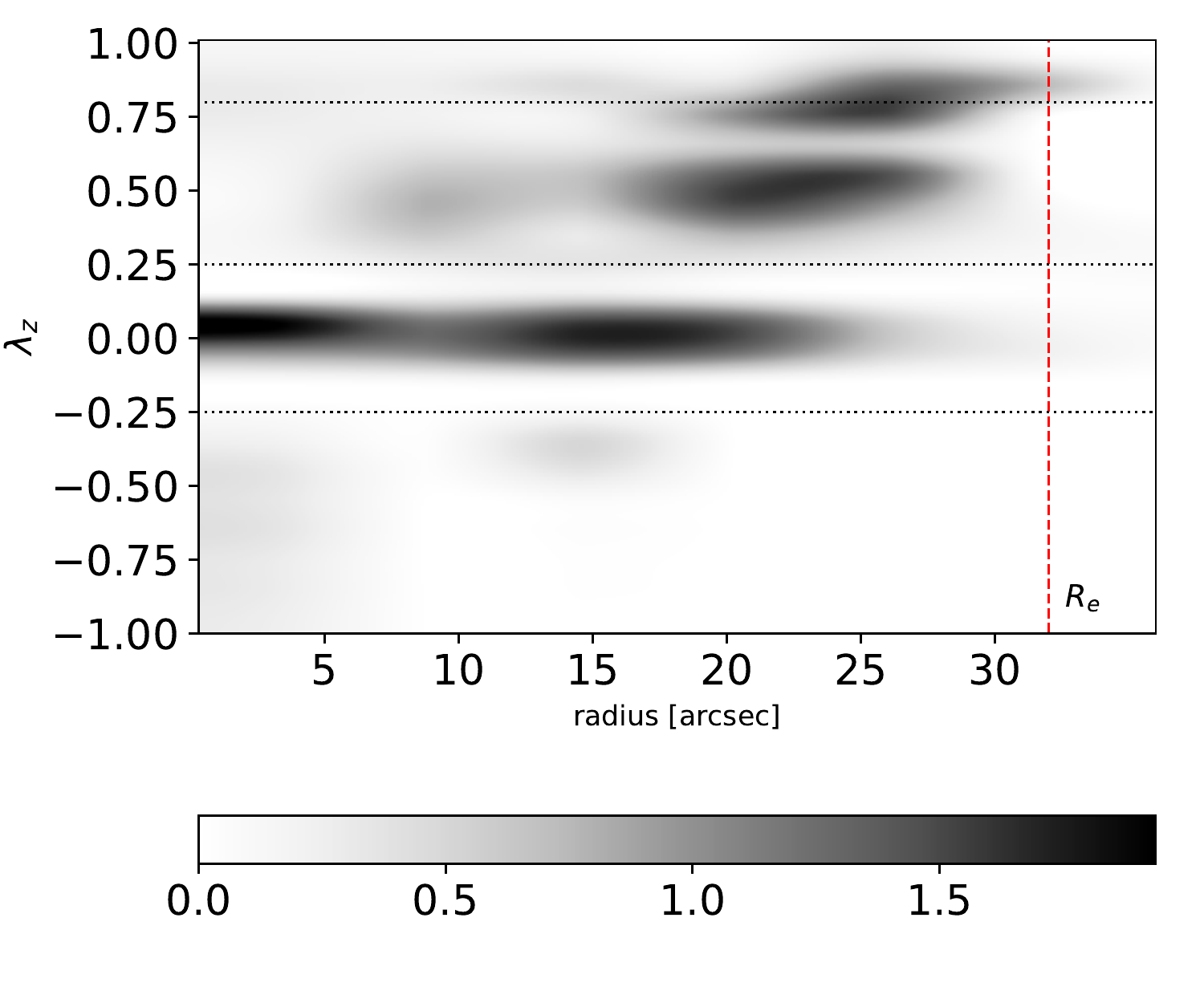}
	\caption{ Orbit distribution in the phase space of circularity $\lambda_{z}$ versus intrinsic radius $r$, derived from the best-fitting orbit-superposition model.  Grey-scale colors indicate the orbital density in phase space and horizontal dashed lines represent the boundaries to separate orbits into 4 classes of  cold ($0.8 \leq  \lambda_{z} \leq 1.0$), warm ($0.25 <  \lambda_{z} < 0.8$), hot ($-0.25 \leq \lambda_{z} \leq 0.25$) and CR ($\lambda_{z} < -0.25$). The red dashed vertical line indicates the effective radius ($R_e$) of NGC~7457.}
	\label{fig:schw-phas}
\end{figure}

\subsection{Reconstruction of the orbital components}

Taken the characteristics and the probability density of each orbit contributing in the best-fitting model, $\rho(\lambda_{z},r)$, in Fig.~\ref{fig:schw-phas} we present the orbits distributions of the best-fitting model for NGC~7457 in the phase space of circularity $\lambda_{z}$ versus intrinsic radius $r$. Following the similar approach presented in \citet{zhu2018}, orbits are divided into four classes of: cold ($0.8 \leq  \lambda_{z} \leq 1.0$), warm ($0.25 <  \lambda_{z} < 0.8$), hot ($-0.25 \leq \lambda_{z} \leq 0.25$) and CR with $\lambda_{z} < -0.25$. It is worth noting, these cuts are consistent with the dips recognised in the orbital distribution of the galaxies in the CALIFA sample \citep{sanc2012}, derived by \citet{zhu2018b} spanning a wide range of masses and morphological types. Moreover, the orbital components selected out in this way are consistent with the dynamical decompositions in a variety of simulations \citep[see Fig. 16 of][]{zhu2018c}. By adopting these boundaries, the luminosity fractions ($f_{L}$) of the cold, warm, hot and CR components within 1$R_e$ are 0.07$\pm$ 0.05, 0.43$\pm$ 0.09, 0.46$\pm$ 0.14 and 0.04$\pm$ 0.02, respectively. The uncertainty in the luminosity fraction of each component represents the scatter among the models within 1$\sigma$ confidence. 

As shown in Fig.~\ref{fig:schw-phas}, the phase space is dominated by hot orbits in $r < 20\arcsec$ and warm orbits in the outer regions, while contribution of the cold component is very small. In Fig.~\ref{fig:M_f_NGC7457} we present the orbits fractions of CALIFA galaxies, as function of galaxy's total stellar mass, $M_{*}$, while the corresponding values for NGC~7457 are overlaid. Clearly, the fraction of both hot and cold orbits in our orbital model of NGC~7457 are more similar to that in most massive early-type than a low mass system in CALIFA sample.

\subsection{Morphology of orbital components}
\label{subsec:sch-morph}

\begin{figure}
	\centering
	\includegraphics[width=0.49\textwidth]{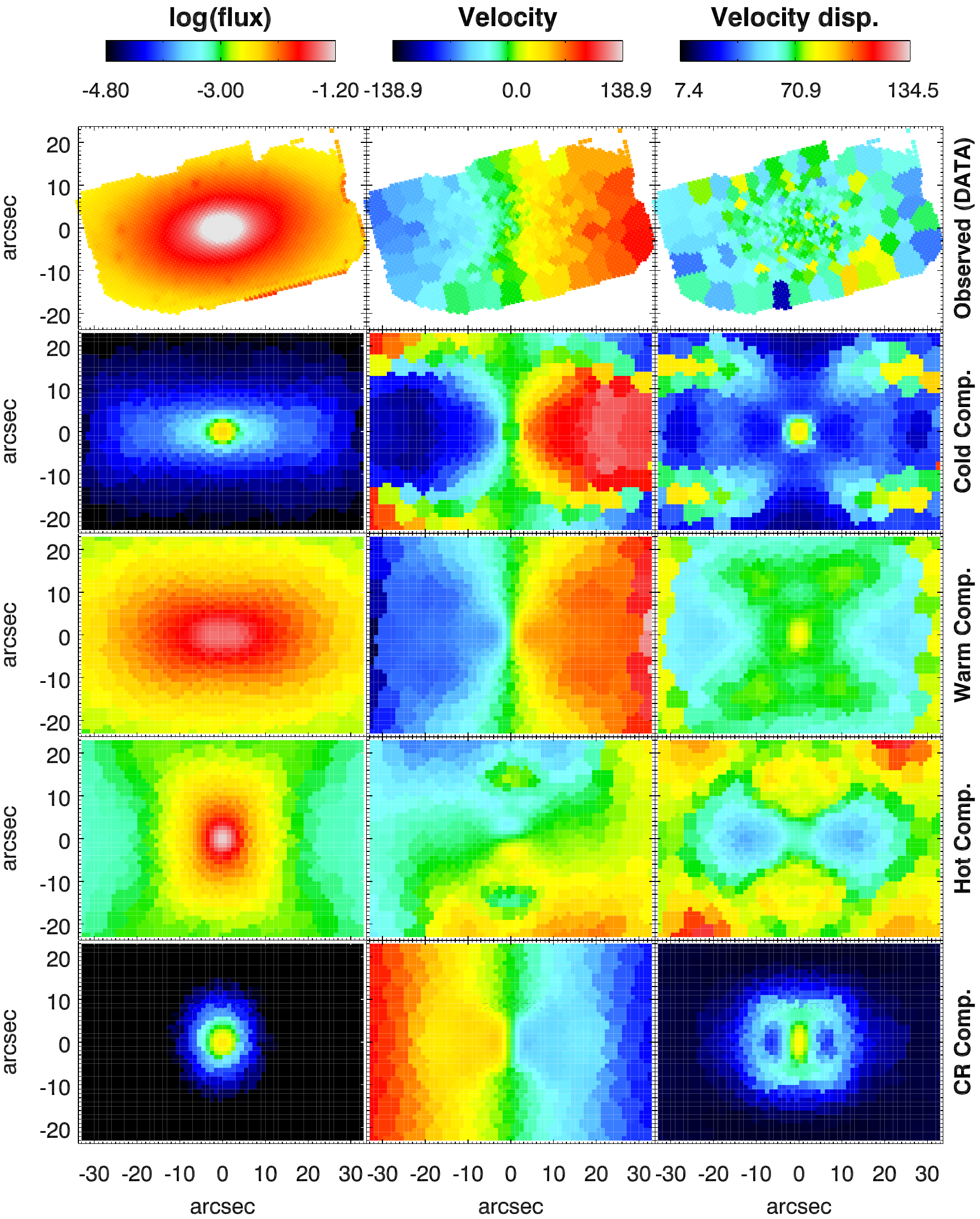}
	\caption{First row: The observed (\sauron) SB and kinematics maps of NGC7457. Second to the forth rows: The reconstructed SB and kinematics maps of the cold, warm, hot, and CR components, extracted from the best-fitting orbit-superposition model of NGC~7457 at the observed inclination ($i=73^{o}$). The SB maps have been weighted according to the contribution of each orbital component to build  and the total luminosity is normalized to unity.}
	\label{fig:sch-recon-obs}
\end{figure}

\begin{figure}
	\centering
	\includegraphics[width=0.46\textwidth]{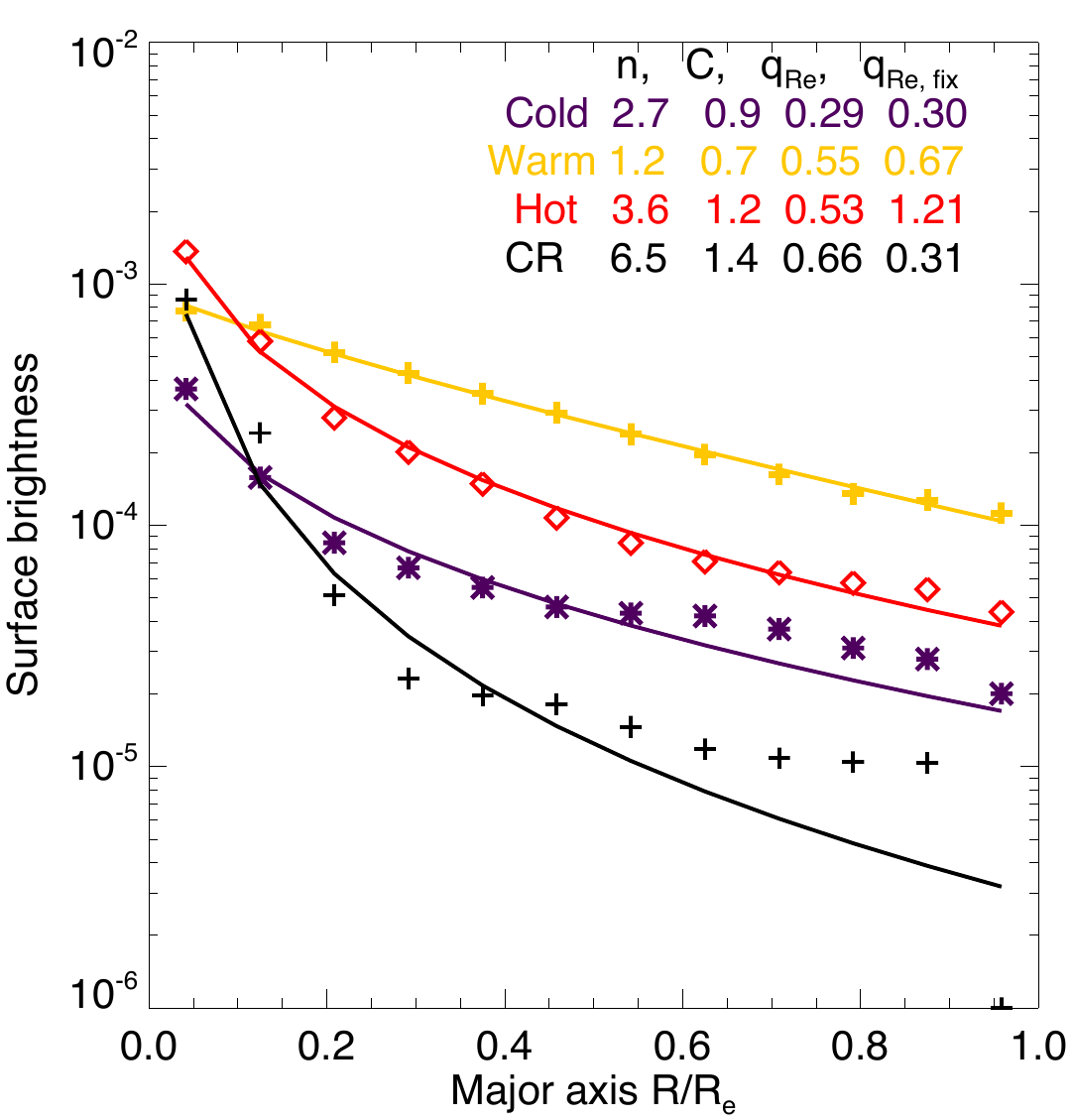}
	\caption{Normalised edge-on surface brightness profiles of the orbital components, derived from the best-fitting model, along the galaxy’s major axis of NGC~7457s. The blue crosses, orange triangles, red diamonds, and black pluses are the averaged SB of the cold, warm, hot, and CR components, respectively.}
	\label{fig:sch-sb}
\end{figure}

\begin{figure}
	\centering
	\includegraphics[width=0.48\textwidth]{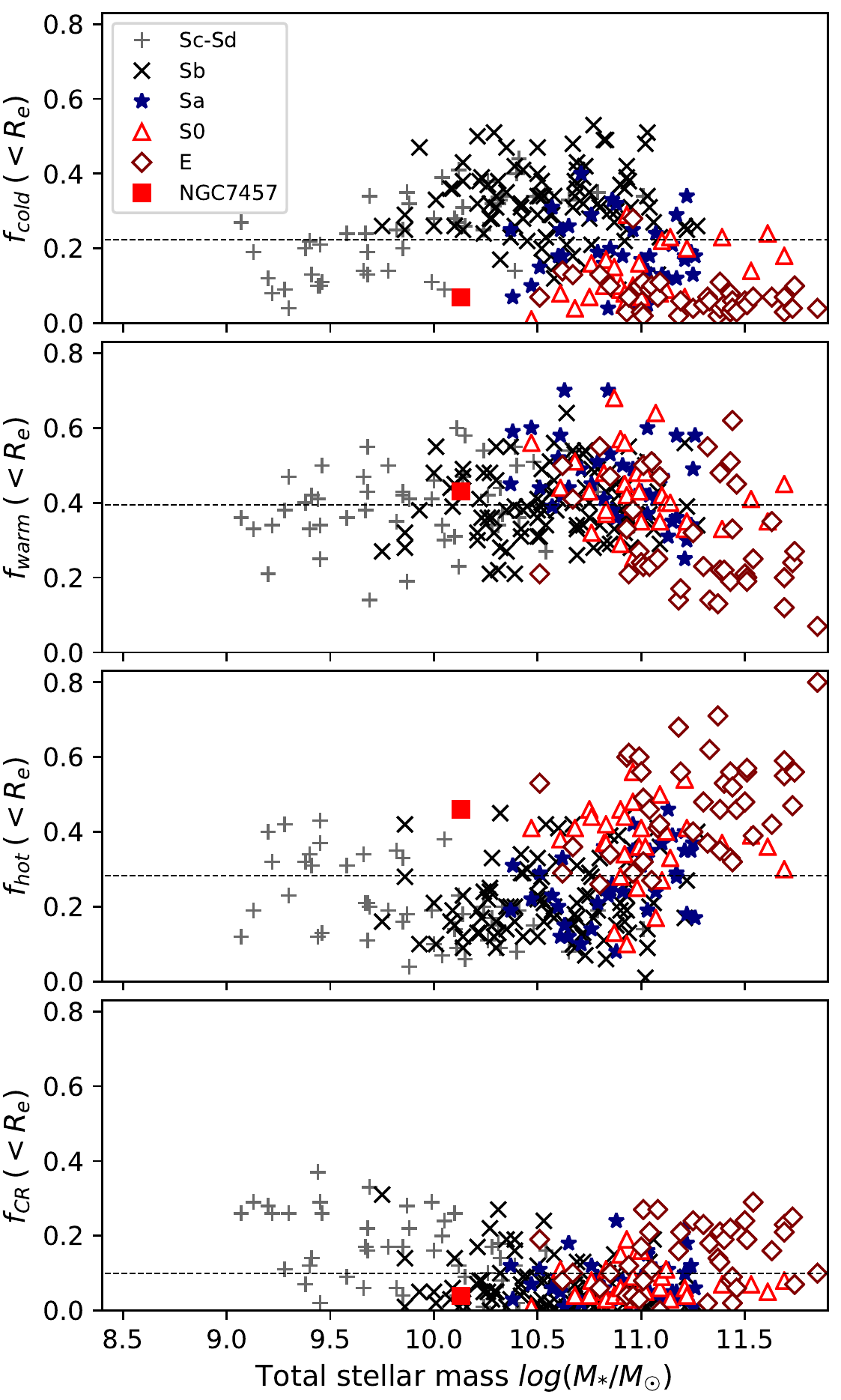}
	\caption{Luminosity fractions of cold, warm, hot, and CR components within $R_{e}$ versus the total stellar mass (obtained from SED fitting with Chabrier IMF) for 300 CALIFA galaxies. The horizontal dashed line in each panel represents the corresponding average value for CALIFA sample over the entire mass range ($10^{9.5} - 10^{11.5} M\sun$). Overlaid (filled red square) are the corresponding values for NGC~7457. }
	\label{fig:M_f_NGC7457}
\end{figure}

In Fig.~\ref{fig:sch-recon-obs} we illustrate the reconstructed SB and stellar kinematics maps of the best fitting model of NGC~7457  for the different orbital components, inclined to the observed inclination of $i=73^{o}$. For the comparison purpose, the observed (\sauron) SB and kinematics maps are also presented (first row). Note that, the SB maps have been weighted according to the contribution of each orbital component to build the final model. To present these results in a more quantitative way, in Fig.~\ref{fig:sch-sb} we illustrate the surface brightness radial profile of these components. Following the approach presented in \citet{zhu2018c}, we have used a set of four parameters namely, S\'ersic index ($n$), concentration ($C$) and two intrinsic flattening parameters $q_{Re}$ and $q_{Re, fix}$, as quantitative descriptors of these profiles \citep[see][for definitions]{zhu2018c}. We have to note here that, for the flattening parameter $q_{Re}$, we adopt the definition that allowing different components having their own major and minor axis, while the $q_{Re, fix}$ is the flattening calculated by fixing the long axis of each component aligned with the major axis of the galaxy.

Interestingly, the cold ($0.8 \leq  \lambda_{z} \leq 1.0$) component of the best fitting model for NGC~7457 has a remarkably high S\'ersic index and concentration ($n$=2.7, $C$=0.9) in the central parts, comparable to the most massive galaxies in \citet{zhu2018b} sample. These results together with the relatively high levels of flattenings ($q_{Re}$ \& $q_{Re,fix}$ $\sim$ 0.3) describe a flattened and near-oblate structure. Clearly, the cold orbits cannot be fit by a single Sersic profile, representing a perturbed disc with stars on nearly circular orbits. Later, in Section \ref{sec:discussion} we will discuss the different possible scenarios that might have lead to the formation of this dynamically cold component. To better understand the geometry of this component, in Fig.~\ref{fig:sch-decom4-i90} and~\ref{fig:sch-decom4-i0} we present the reconstructed SB and kinematics maps of of the different classes of orbits in our model in two edge-on (inclination $i= 90^{o}$) and face-on ($i= 0^{o}$) projections. Warm orbits in our best fitting model remarkably contribute in both inner and outer parts of the surface brightness map. The SB profile of the warm orbits are well described by an exponential profile ($n_{Sersic}=1.2$), constitute a thick disc. The next interesting feature is an elongated (and triaxial) spheroid, representing the SB of hot orbits in our best-fitting model. This orbital component with the highest contribution in the total luminosity of the best-fitting model, demonstrates significantly high level of S\'ersic index ($n$=3.6) and intrinsic flattening ($q_{Re,fix}$= 1.21). The surface brightness of the hot component fairly matches the photometric bulge, with comparable scale length ($\sim 10\arcsec$). The reconstructed LOSVD map of this component shows clear rotation around the major photometric axis of this galaxy, known as "prolate rotation" \citep[see][for more details]{tsat2017}. The CR component of the best fitting model is highly concentrated ($C$=1.4) in the central regions ($r < 2\farcs5$). The scale lenght of the CR component matches quite well the measured diameter of the KDC, observed in the centre of NGC~7457 \citep[e.g.][]{emse2004}. However, It should be stressed that, as noted in \citet{zhu2018}, the dynamically decomposed components are not necessary to match the morphologically decomposed ones. Another important caveat is that, the distribution of warm orbits in NGC~7457 ($0.25 <  \lambda_{z} < 0.8$) shows a bimodal pattern with a dip at $\lambda_{z} \approx0.7$. Hence, it would be tempting to make the cut between cold and warm components not at $\lambda_{z}=0.8$, but instead at the dip around $\lambda_{z}=0.7$. We tested this and reassuringly concluded that, despite a remarkable increase in the fraction of cold orbits, the shape of the SB and kinematics of neither cold nor warm component has changed significantly. Therefore, our results are robust against changing the boundaries to define the warm and cold components. We will come back to this issue in Section \ref{subsec:disc_scenario}, where we discuss the possible scenarios for the formation of the thick disc in NGC~7457.


 \begin{figure*}
  	\label{fig:sch-decom4}
 	\captionsetup[subfigure]{labelformat=empty}
 	\centering
 	\begin{subfigure}{0.5\textwidth}
 		\centering
 		\caption{inclination $i=90^{o}$ (edge-on)}
 		\includegraphics[width=1.0\textwidth]{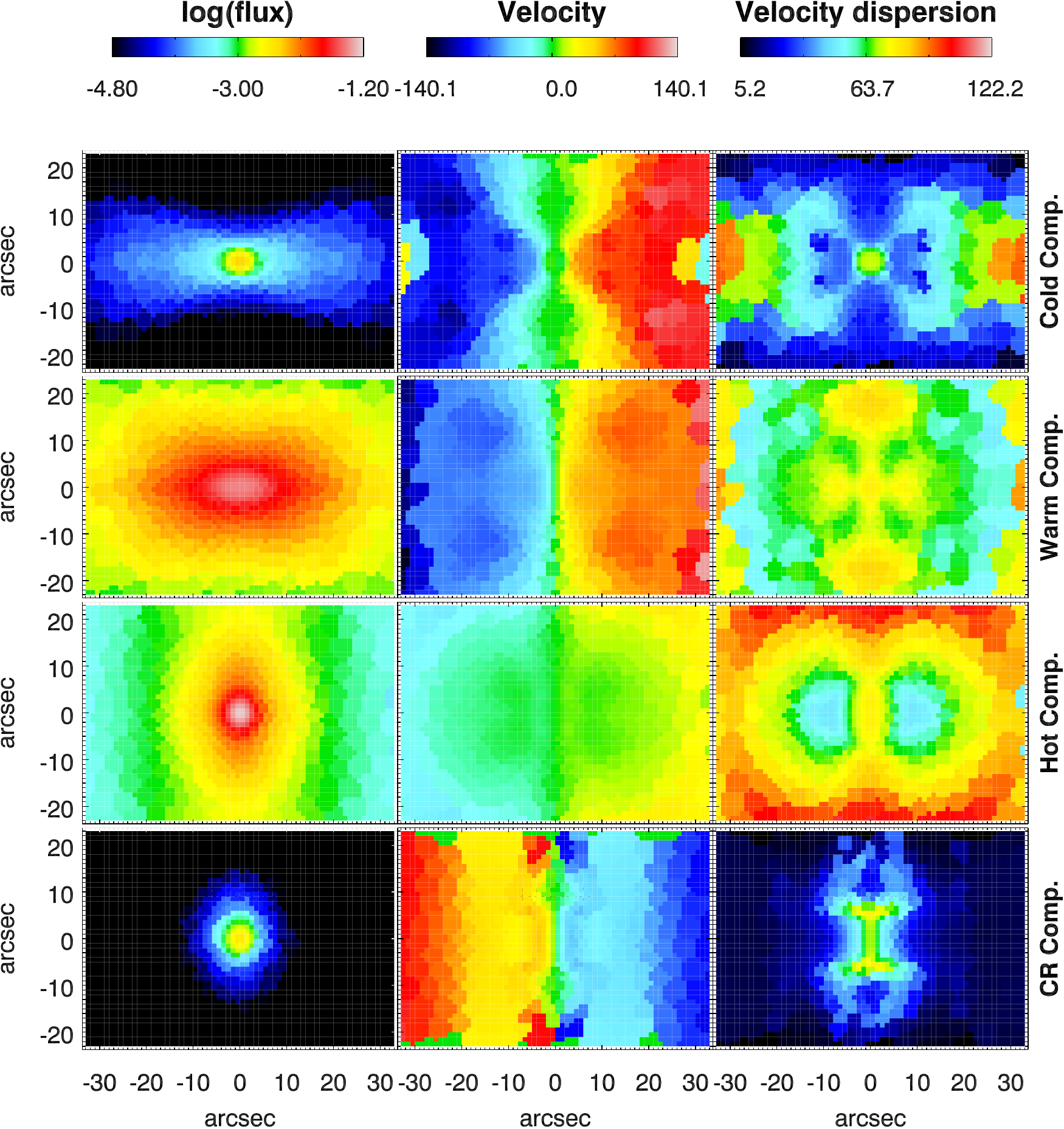}
 		\label{fig:sch-decom4-i90}
 	\end{subfigure}%
 	\begin{subfigure}{0.5\textwidth}
 		\centering
 		\caption{inclination $i=0^{o}$ (face-on)}
 		\includegraphics[width=1.0\textwidth]{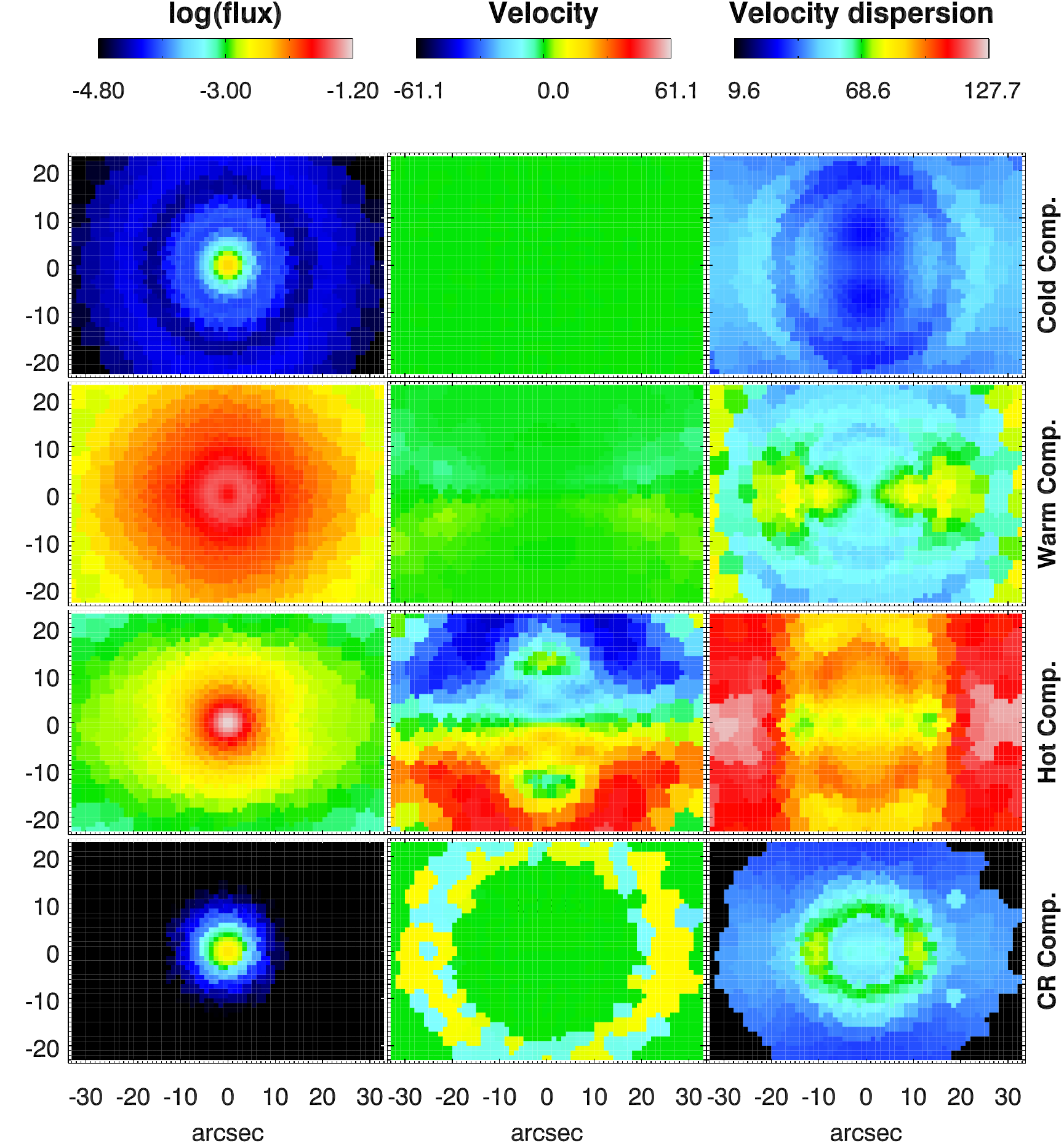}
 		\label{fig:sch-decom4-i0}
 	\end{subfigure}
 	\centering
 	\centering
 	\caption{The reconstructed SB and kinematics maps of the cold, warm, hot, and CR components, derived from the best-fitting orbit-superposition model in edge-on (left panel) and face-on (right panel) projections. The SB maps have been weighted according to the contribution of each orbital component to build the final model and the total luminosity is normalized to unity.}
 \end{figure*}


\begin{figure}
	\centering
	\includegraphics[width=0.5\textwidth]{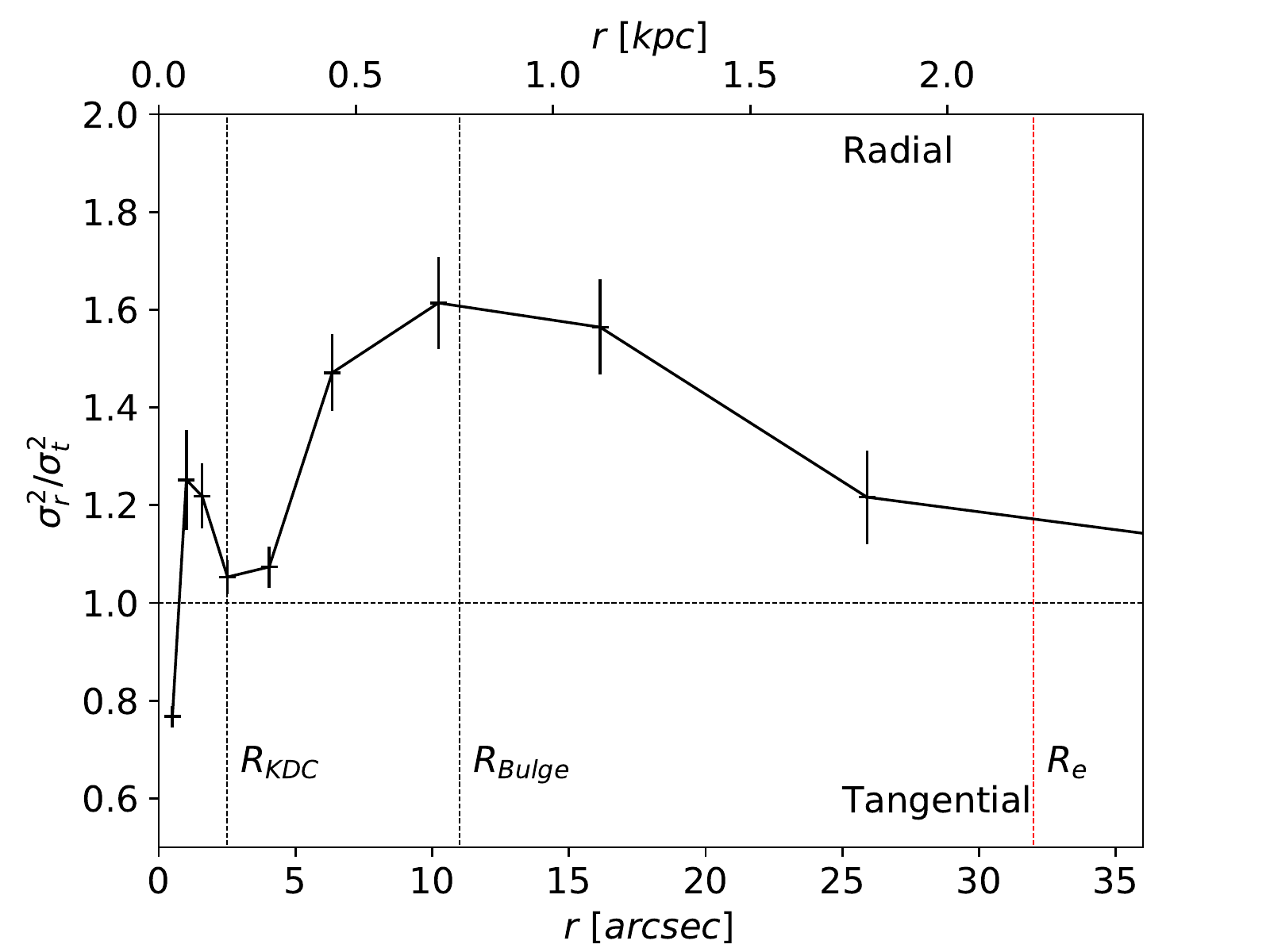}
	\caption{The velocity anisotropy profiles $\sigma^{2}_{r} / \sigma^{2}_{t}$ of the best-fitting model for NGC~7457 as a function of the intrinsic radius, $r$. Errors represent the scatters among the models within 1$\sigma$ confidence. The two black vertical dashed lines have been drawn to show the rough extents (along the major axis) of the KDC- and bulge-dominated regions (photometric bulge) and the red vertical line indicates the effective radius of NGC~7457.}
	\label{fig:anisotropy}
\end{figure}

\begin{figure}
	\centering
	\includegraphics[width=0.5\textwidth]{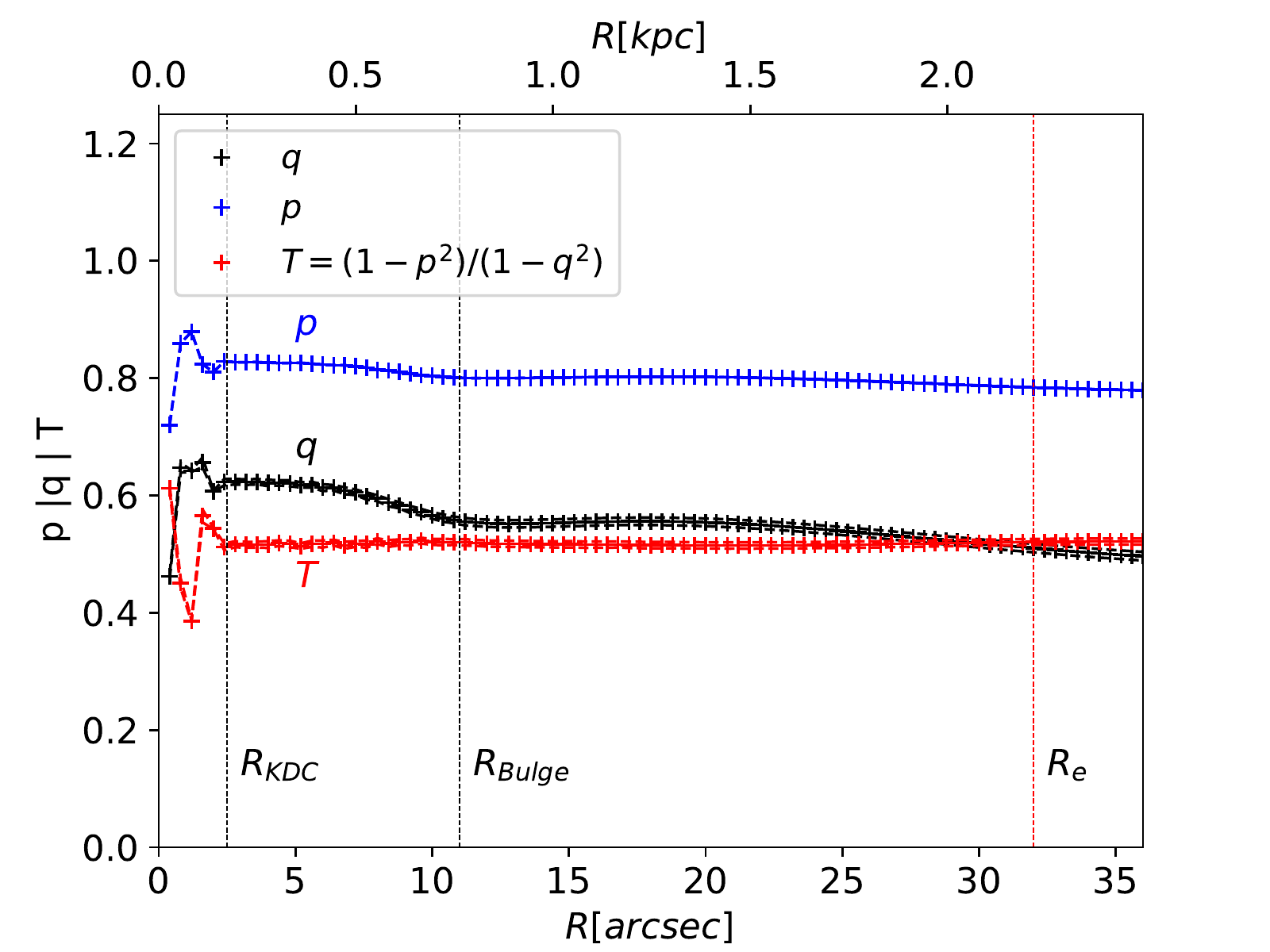}
	\caption{$q$, $p$ and Triaxiality parameter $T$, as a function of distance from the centre of best-fitting model for NGC~7457. The two vertical dashed lines indicate the rough extents (along the major axis) of the KDC- and bulge-dominated regions. The effective radius ($R_e$) is marked with the red dashed vertical line.}
	\label{fig:triaxial_qpt}
\end{figure}

\section{Discussion}
\label{sec:discussion}

In the preceding sections of this paper, we investigated the stellar kinematics and populations of NGC~7457 within $\sim1R_e$, allowed by our
\sauron\ data. Using a triaxial implementation of Schwarzschild’s (1979) orbit superposition method, we constructed a realistic triaxial dynamical models for this galaxy and dynamically decomposed the best-fitting model into four orbital substructures.  Our r--band M/L of 2.29 $_{-0.33}^{+0.15}$ inferred from the best-fitting Schwarzschild modelling
is slightly lower than 2.8 reported in \citet{capp2013}, from stellar population modelling, assuming Salpeter IMF \citep{salp1955}, though it is consistent within $2\sigma$ errors. This difference most-likely arise from different IMFs, assumed in these two studies. 
 The inclination angle $\theta$, derived from the intrinsic shape parameters is $\sim$ 73$^{o}$ in excellent agreement with inclination of 74$^{o}$, reported in \citet{capp2013b}. 

In order to investigate the dynamical behaviour of this galaxy, in Fig.~\ref{fig:anisotropy} we present the velocity anisotropy profiles $\sigma_{r}^{2}/\sigma_{t}^{2}$ as a function of the intrinsic radius $r$ for the best fitting-model, where $\sigma_t \equiv \sqrt{(\sigma_{\theta}^{2}+\sigma_{\phi}^{2})/2}$ and $\sigma_{r}$, $\sigma_{\theta}$ and $\sigma_{\phi}$ represents the radial, azimuthal angular and polar angular velocity dispersion in a spherical coordinates.The two vertical dashed lines in this figure indicate the rough extents of the KDC- and bulge-dominated regions, along the major axis. According to this profile, NGC~7457 seems consistent with isotropy in the inner regions and mild radially anisotropic towards larger radii. To provide an estimate of the intrinsic (3D) outer shape of the best-fitting model, in Fig.~\ref{fig:triaxial_qpt}, we present the radial profile of the $q$ and $p$ parameters introduced in Section~\ref{subsec:sch-dyn-modl}, and the triaxiality parameter, $T$ (defined as $T \equiv 1-p^2/1-q^2$) along the major axis. Comparing these profiles with the orbits distributions in the phase space (Fig.~\ref{fig:schw-phas}), one can argue that, purely on the basis of the orbital configurations, the overall structure of this galaxy is a combination of a triaxial and dynamically hot bulge, surrounded by a kinematically warm and isotropic thick disc. However, as described in Section~\ref{subsec:sch-morph}, the reconstructed surface brightness of different orbital components show a variety of complex features.

\subsection{NGC~7457's disc: born hot or dynamically heated?}

\label{subsec:disc_scenario}

 In the absence of a dominant thin disc, the outer part of this galaxy is dominated by a dynamically-warm component resembling the 'thick disc' in spiral galaxies. This is in favour of those formation scenarios, in which i) disc stars were dynamically hot at birth \citep[e.g.][]{bird2013} or ii) disc stars born in a very thin layer of gas with cold orbits and observed thickening have appeared more recently \citep[e.g.][]{merr2001}. 

In the first scenario (born hot thick disc), discs being formed originally thick in situ at high redshift by the merger of gas-rich protogalactic fragments and later thin discs may be formed \citep{broo2007,bour2009,come2014,come2016}.
 However, as noted in \citet{mcde2015} sometimes lenticular galaxies never acquired thin discs and only possess thick one \citep[see][for more details]{kasp2016}. This scenario should produce old and low metallicity stellar population in the thick disc, which is in contrast to the generally young population of stars, observed over the entire \sauron\ field of NGC~7457. 

On the other hand, in a disc heating scenario, stars were born dynamically cold in a primordially thin discs and can get dynamically heated and form a thick disc through different mechanisms e.g. satellite flybys \citep[see][]{quin1993,qu2011a,qu2011b}, disturbances by satellite galaxies or mergers \citep[e.g.][]{gers2012} and secular thin disc flaring \citep{scho2009b}. In the latter mechanism, the thick disc being secularly formed through the heating processes induced by disc over-densities \citep{vill1985} or dark halo objects and globular clusters or through the radial migration mechanisms \citep{scho2009b}. Given that, NGC~7457 is a relatively low-mass and dynamically young galaxy, any long-term secular scenario is unlikely. Moreover, there is no observational evidence in favour of possible progenitor of thick disc in our kinematical, chemical properties and dynamical modelling of this galaxy.

\begin{figure}
	\centering
	\includegraphics[width=0.5\textwidth]{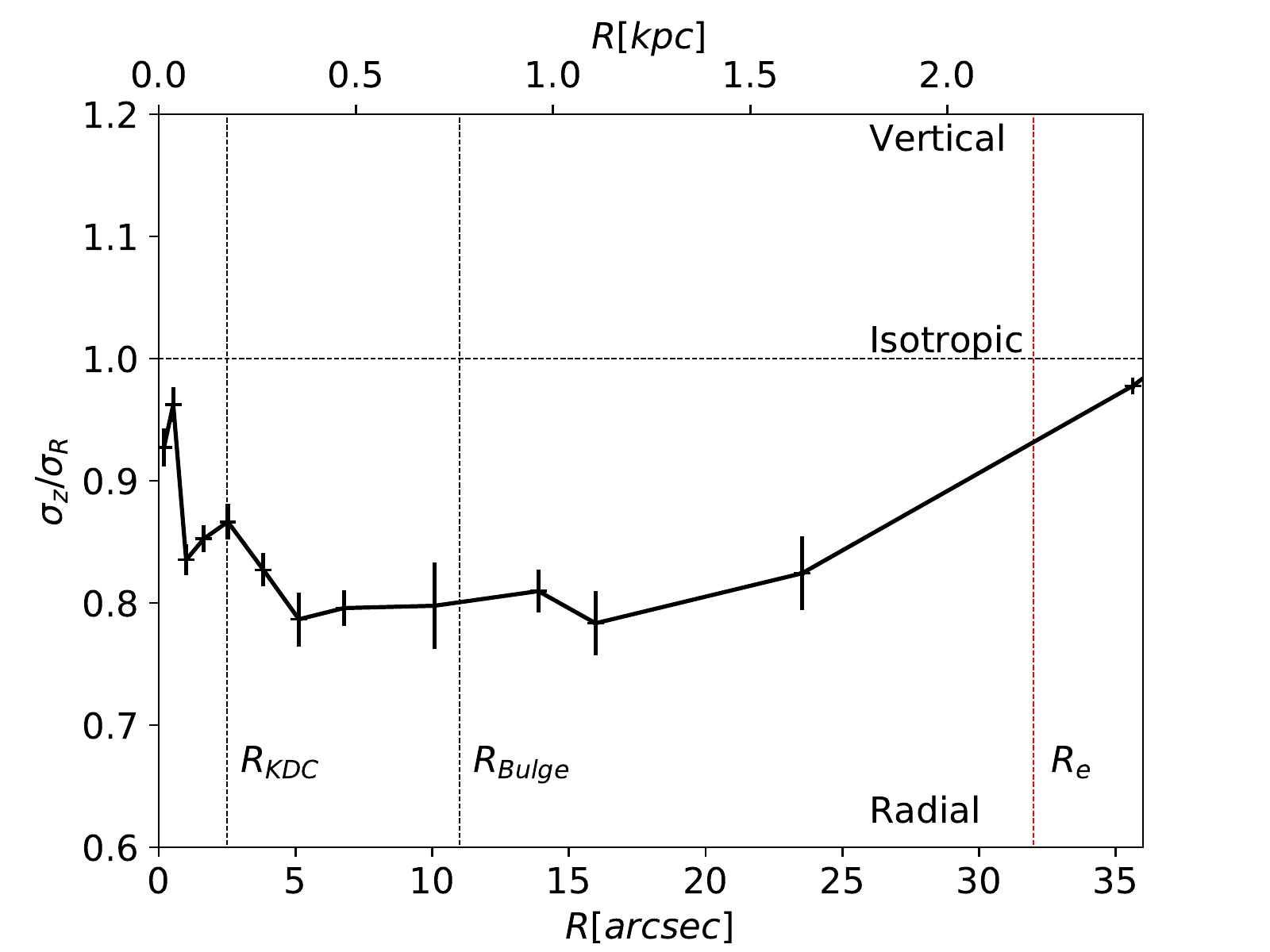}
	\caption{The stellar velocity ellipsoid, $\sigma_{z} / \sigma_{R}$ as a function of radius ($R$) on the disc plane. The two black vertical dashed lines indicate the rough extents (along the major axis) of the KDC- and bulge-dominated regions and the effective radius ($R_e$) is marked with the red vertical dashed line. The error bars represent the scatters among models within 1$\sigma$ confidence intervals.}
	\label{fig:betaz_var}
\end{figure}

The values of $\sigma_{z}/\sigma_{R}$, (where $\sigma_{z}$ and $\sigma_{R}$ are the vertical and radial velocity dispersions in a cylindrical coordinates) have been widely used in literature as a discriminant of different disc heating agents in the spiral galaxies \citep[e.g.][]{merr2001} and the radial profile of this ratio is found to be a good indicator of the initial inclination of the decaying satellite \citep{vill2008}. However a possible caveat is that, as noted by \citet{pinn2018}, the stellar velocity ellipsoid is very sensitive to the condition of possibly different heating processes that a galaxy may experience during its life-time \citep[refer to][for more details]{pinn2018}. In Fig.~\ref{fig:betaz_var} we present the $\sigma_{z}/\sigma_{R}$ for our best-fitting model along the radius $R$ on the disc plane. As it shows, the ratio is close to unit along the radius, however it is moderately lower within 0.5$R_{e}$ and slightly increase afterwards and reaches its maximum at $\sim$ $R$=1$R_{e}$. 

\citet{vill2008} presents simulations of the heating of a disc galaxy by a single minor merger (e.g. mass ratios
of 1:5 to 1:10). They probed several configurations of the progenitor, including three different initial orbital inclinations of the satellites in both prograde and retrograde directions with respect to the rotation of the host disc and different mass scales. They studied the trend of the ratio $\sigma_{z}/\sigma_{R}$ of the thick disc component as a function of radius for different prograde experiments and argued that the radial profile of the stellar velocity ellipsoid, $\sigma_{z}/\sigma_{R}$  in the final thick disc can be considered as a good indicator of the initial inclination of the decaying satellite. They found that satellite interactions with 10--20 percent mass of the host galaxy can efficiently heat up the pre-existing disc in host galaxy to give rise to a thick disc, while its scale-height increases in proportion to the inclination of the encounter. They also argued that satellite encounters with lower initial inclinations are more efficient in introducing asymmetric drifts in the mean rotational velocity of the final thick disc;  So that, encounters at higher inclinations can decrease the vertical gradients in the line-of-sight velocity distribution (LOSVD). Comparing the velocity ellipsoid ($\sigma_{z}/\sigma_{R}$) of the best-fitting model for NGC~7457 and those predicted by this simulation study, we suggest that the thick disc in NGC~7457 is most likely a dynamically heated structure, formed through the interactions and accretion of satellite(s) with near-polar initial inclinations. We suggest that this scenario can also explain the elongation of hot orbits, resembling the photometric bulge of NGC~7457, depending on the time and the conditions in which these encounters occur. 

NGC~7457 is located in a relatively low density environment \citep{ludw2012} but still accretion or minor merger(s) in the past seem inevitable. Assuming a single satellite accretion as a candidate agent for the dynamical heating of NGC~7457 disc, the presence of a dynamically warm and extended disc could be well explained as the result of a satellite accretion with an intermediate to high initial inclination. Moreover, as noted in Section \ref{subsec:sch-morph}, cold component of the best fitting model for NGC~7457 represents a perturbed disc-like structure. It might be well the result of a (prolate) merger perturbing the original (close to) exponential thin disc. It clearly supports our proposed near-polar merger scenario in which the circularity ($\lambda_{z}$) distribution of stars in the original thin disc shifts to lower regime, through the disc heating process.

The best-fitting dynamical model of NGC~7457 clearly shows a counter-rotating orbital component in the central regions of this galaxy. This feature has been observationally confirmed as an existence of a KDC \citep[e.g.][]{silc2002,emse2004}. This KDC is geometrically matched to the central chemically distinct young, metal-rich stellar populations presented in Section~\ref{sec:ssp} for the centre of NGC~7457. However, as noted in Section~\ref{subsec:sch-morph} our analysis of the orbital structure of NGC~7457 suggest a complex mixing of different orbital families and structures in the very centre. Therefore we could not blindly assign our populations properties to these dynamical substructures. We should also take into account that, this galaxy is known to host AGN \citep{gebh2003}, that clearly makes the evolution path of this galaxy much more complex. 

Assuming the young stellar populations belong to the counter-rotating core of the gas-poor NGC~7457 \citep{pele1999,welc2003,sage2006}, gas is required to be present during the formation event. It is in favour of those formation scenario that involves a gas rich merger with retrograde orbital configurations. The spatial distribution of the counter-rotating component depends on the initial angular momentum of the acquired gas \citep[see][for more details]{pizz2018}, such a way that it reaches to the centre with the opposite spin, being compressed and consumed by star formation to form a counterrotating stellar component \citep{silc2011}. An alternative process, without resorting to gas dynamics, would be the creation of a kinematically decoupled core through the retrograde stellar spiral-spiral mergers \citep{balc1998}. In this schema, the central bulges transport orbital angular momentum inward to the centre of the remnant, while the outer parts keep the spin signature of the precursor discs. The resulting KDC in this schema is expected to be metal-rich with traces of a young population born from gas provided by the merger. However, we should consider the possibility that a single merger had in addition to the polar angular momentum also part of it total orbital angular momentum opposite to the main host disc-plane and therefore, the gas that the merger brought in settle, also contribute in forming a counter-rotating component in the centre. In this way, one can explain the formation of both the thick disc and the KDC at once.

\subsection{NGC~7457's bulge}

Hot component in our best-fitting model presents an elongated spheroid, representing the photometrically-identified bulge. This orbital component shows high level of triaxiality and the reconstructed velocity map of this component shows rotation around its major apparent axis (prolate rotation). This feature has been observed in few massive early-type galaxies, mostly belong to galaxy groups or clusters \citep{davi2001,emse2004,kraj2011,emse2011}. Given the rarity of observations for this dynamical feature, its origin is still under debate and not well understood. \citet{tsat2017} investigated a possible merger origin of prolate rotation by studying the kinematics of simulated early-type massive galaxies formed in N-body merger of progenitor spiral galaxies in polar orbits. They also reported ten early-type galaxies from the CALIFA Survey show observational evidence for prolate rotation. Given that they found no evidence for oblate rotation in these galaxies, they suggest a gas-poor merger origin for these systems. As noted earlier, comparing to the low mass galaxies in the CALIFA sample ($M_* < 10^{10} M_\odot\ $), the hot component of NGC~7457 has too large $n$ and $C$, as well as too large $q$. In fact, as noted in \citet{mend2018} for CALIFA galaxies at this mass range, bulge contribution is rather small ($B/T < 0.2$) and hot components are usually disc-like, not as bulge-like as NGC~7457.
Moreover, no prolate rotation has been reported in this mass range of the CALIFA galaxies, all late-type systems. We should note that, the CALIFA data quality of these low-mass galaxies are not good enough and these systems are mostly edge-on, thus it is very likely if there is actually similarly kinematic twin as NGC~7457, it could not be well resolved, and may not be fitted by the model. On the other hand, the prolate rotation in the bulge of NGC~7457 coexists with oblate rotation of the cold and warm components, while we do not detect such behaviour in CALIFA sample, therefore, it is not straightforward to interpret the prolate rotation of hot component in NGC~7457 as a merger product, as \citet{tsat2017}. 

We found several studies in the literature suggesting a merger driven mechanism to form the bulge of NGC~7457 from different point of views. \citet{chom2008}, used archival HST/WFPC2 images and Keck spectroscopy to investigate the stellar population of globular cluster (GCs) in NGC 7457 and concluded that a gas-rich major merger is needed to explain high frequency of old and metal-rich GCs in this galaxy. Later, similar conclusion made by \citet{harg2011}, who using the same approach and also reported a noticeably elliptical spatial distribution for old and metal rich GCs in NGC~7457. Recently, \citet{zana2018} presented a detailed chromodynamical analysis of NGC~7457 GCs and argued that, comparing the age of GCs ($\sim$ 3-7 Gyr) and estimated halo assembly epoch of NGC~7457 \citep[$\sim$ 12.3 Gyr;][]{alab2017} a gas-rich major merger is not the right mechanism to form these structures but still accretion and interaction of gas-rich satellite seem inevitable. A similar conclusion has been made by \citet{bell2017}, who studied the stellar kinematics of the lowest stellar mass lenticular galaxies in SLUGGS survey \citep{brod2014}, including NGC~7457 and by comparing the stellar kinematics of their sample with with those from various simulated galaxies, concluded that all low mass lenticular galaxies in their sample have not experienced major mergers since $z\sim1$, but they have probably experienced accretion through gas-rich minor mergers.

As noted in \citet{ludw2012}, NGC~7457 resides in one the 3 subgroups of the young filamentary group NGC~7331. NGC~7457 is accompanied by an asymmetric and spindle-shape dwarf S0/dE, UGC~12311 with the angular separation of 5.7\arcmin\ ($\Delta V < 80\; km/s $). NGC~7457 subsystem represents the most dynamically evolved substructure of the NGC~7331 group \citep{ludw2012}. This is in favour of the merger-driven evolution history we proposed to explain the observed properties and the orbital distribution in NGC~7457. For instance, as shown in \citet{balc1998}, a disc-disc merger with mass ratio 3:1 or larger (intermediate mass ratio merger) does not destroy the orbital structure of the pre-existing disc, rather it heats it up leading to something resembling an S0, as a result of orbital redistribution.

It seems necessary to look in detail for evidences in the NGC~7457's environment, tracing such possible interactions. It is worth referring to the work of \citet{duc2015}, who adopted a new observing strategy and data-reduction techniques to explore the diffuse low surface brightness light around nearby galaxies. Interestingly, they reported a stellar stream similar to a tidally disrupted satellite emanating east direction of NGC~7457 \citep[see Fig. 10 of][]{duc2015}. However, given the similarity of this filamentary tail to those present all over the field of this galaxy in a zoom-out view, they suggested that this is most likely due to Galactic cirrus emission. 

\subsection{Cylindrical rotation in the absence of an strong bar}

 Given that there is no reliable evidence in favour of an strong bar-like structure in NGC~7457, we propose a different scenario than a bar-driven kinematics to explain the vertically aligned LOSVD, dubbed ‘cylindrical rotation’, within the photometric bulge region of this galaxy. As noted in some N--body and numerical simulations of satellite encounters, such interactions with an oblique impact angle of the satellite can disrupt the radial motion pattern of stars in host galaxies and boost the vertical motion of disc particles \citep[e.g.][]{bure1999, vill2008}. Moreover, presence of a counter-rotating component in the centre of NGC~7457, dynamically confirmed in this study, can naturally help to diminish the vertical gradient in the LOSVD. On top of that, as discussed earlier, the reconstructed velocity map of the hot component in the best-fitting model of this galaxy shows rotation around its major apparent axis. Therefore, we suggest that, the observed high level of cylindrical rotation in bulge of NGC~7457 as well as unusually low velocity dispersion of this galaxy could be the outcome of the mixing of (at least) these three mechanisms, all with external origin, with strong evidence against the presence of a bar and hence ruling out the internal origin for the pattern of cylindrical rotation in NGC~7457. 
 
 Could NGC~7457 have had a bar which imprinted the cylindrical rotation and was later destroyed, so that it is undetected today? While the presence of a luminous nucleus in NGC~7457 could contribute to bar destruction \citep{atha2003,bure2005,atha2005}, recent simulations indicate that bar destruction requires unnaturally-high central mass concentrations, as compared to those in real galaxies, and, certainly, as compared to the central mass in NGC~7457 \citep[refer to][for more details]{atha2013}. Therefore, this option seems unlikely. Putting all these pieces together, we suggest a merger-driven evolutionary path to explain the observed photometric, kinematics and populations properties of NGC~7457.
 
\section{Summary and Conclusions}
\label{sec:conclusion}

We constructed traxial Schwarzschild orbit-based models of NGC~7457 and presented a detailed orbit-modelling analysis of this peculiar lenticular galaxy, with stellar kinematics and populations obtained from \sauron\ IFU observations out to $\sim1R_{e}$ of this galaxy. We reconstructed the surface brightness and kinematics maps of different orbital components, obtained from the best-fitting model and investigated possible mechanisms to explain peculiarities reported for this galaxy in previous studies. We succeeded in putting together all puzzle pieces of observed photometric, kinematics and populations of NGC~7457 to better understand the evolutionary path of this peculiar galaxy. The main results of this study are:

\begin{itemize}
\item Our kinematics analyses of NGC~7457, using \sauron\ IFU data exhibit high level of cylindrical rotation ($m_{cyl} =0.83 \pm$ 0.06) within the bulge dominated regions. We also confirm the previously reported low velocity dispersion ($<\sigma_{1Re}>=63.5\pm3.5$ \ensuremath{\mathrm{kms^{-1}}}) for this galaxy. 

\item In agreement with the recent detailed photometric studies of NGC~7457 \citep[e.g.][]{erwi2015}, we find no evidence in favour of a bar-like structure in our kinematics and stellar population analyses of this galaxy. 

\item Our stellar population analysis of this galaxy confirm that, in agreement with previous studies, the bulge of NGC 7457 is dominated by 4-5 Gyr stellar population with nearly solar metallicities and over-solar [Mg/Fe] ratios. We also confirm a chemically distinct young stellar nucleus with remarkably high metallicities in the very central regions of this galaxy ($r< 2\farcs5$).

\item The stellar orbit distribution of the best-fit Schwarzschild model of NGC~7457 within $1R_e$ is dominated by warm ($f_{Warm}$= $43\pm 9\%$) and hot orbits ($f_{Hot}$= $46\pm 14\%$), while the contribution of cold and counter-rotating components is small ($f_{Cold+CR}<11 \%$). 

\item The best-fitting dynamical model of NGC~7457 can successfully recover the observationally confirmed KDC in the very centre of this galaxy. This orbital structure is geometrically matched to the chemically distinct young and metal-rich central component of NGC~7457. 

\item  In the absence of a dominant cold component in our triaxial orbit-based model of NGC~7457, the outer part of our model is dominated by warm orbits. The surface brightness of this orbital component robustly represent an exponential thick disc, matches the photometric disc. Comparing the trend of the velocity ellipsoid ($\sigma_{z}/\sigma_{R}$) of the best-fitting model for NGC~7457 and those predicted by the simulations of the heating of a disc galaxy by a single relatively massive merger \citep{vill2008}, and given the relatively young and metal-rich populations of stars in NGC~7457, we suggest that the thick disc is most likely a dynamically heated structure, formed through the interactions of satellite(s) with near-polar initial inclination. 

\item The highest level of contribution in the total luminosity of the best-fitting model of NGC~7457 belongs to the hot orbits. The surface brightness of this orbital component matches fairly well the photometric bulge. The reconstructed LOSVD map of this component shows clear rotation around the major photometric axis of the galaxy (prolate rotation). While the formation mechanism of  the NGC~7457 bulge is still not well understood, we have gathered all the published studies of NGC~7457 bulge and suggest a merger-driven process as the most plausible scenario to explain the observed and dynamically-modelled properties of NGC~7457 bulge. 

\end{itemize}

We suggest, both high level of cylindrical rotation and unusually low velocity dispersions reported for the NGC~7457 bulge could have external origins.  Our suggestion is that, possible satellite interactions with an oblique impact angle (near-polar) can disrupt the radial motion pattern and consequently the NGC~7457 LOSVD. This feature is boosted in the bulge dominated regions, where the counter-rotating component and prolate rotation of stars on hot orbits helps to diminish the vertical gradient of the LOSVD and construct the observed cylindrical rotation pattern. Therefore, we argue that, NGC~7457 provides the first direct example of cylindrical rotation in bulges, driven by external agents and not by bar-driven secular evolution. 

\section*{Acknowledgments}
AM and JALA were supported by the Spanish MINECO grants AYA2017-83204-P and AYA2013-43188-P. LZ acknowledges support from Shanghai Astronomical Observatory, Chinese Academy of Sciences under grant NO.Y895201009. The authors acknowledge support from the Spanish Ministry of Economy and Competitiveness (MINECO) through grants AYA2009-11137, AYA2016-77237-C3-1-P and AYA2014-58308-P. GvdV acknowledges funding from the European Research Council (ERC) under the European Union's Horizon 2020 research and innovation programme under grant agreement No 724857 (Consolidator Grant ArcheoDyn). Funding for SDSS-III has been provided by the Alfred P. Sloan Foundation, the Participating Institutions, the National 
Science Foundation, and the U.S. Department of Energy Office of Science. The 
SDSS-III web site is http://www.sdss3.org/.

\bibliographystyle{mn2e}
\bibliography{ngc7457}

\label{lastpage}

\end{document}